\documentclass[aps,prl,preprint,groupedaddress]{revtex4-1}
\usepackage{color}
\usepackage{graphicx}
\usepackage{amsmath}
\usepackage{ulem}

\begin{document}


\title{Dissipative properties of relativistic two-dimensional gases}


\author{A. L. Garc\'ia-Perciante}
\email[]{algarcia@correo.cua.uam.mx}

\author{A. R. M\'endez}
\email[]{amendez@correo.cua.uam.mx}
\affiliation{ Universidad Aut\'onoma Metropolitana - Cuajimalpa \\
Departamento de Matem\'aticas Aplicadas y Sistemas\\
           Av. Vasco de Quiroga 4871, M\'exico D.F., 05348, M\'exico.\\}


\date{\today}

\begin{abstract}
The constitutive equations for the heat flux and the Navier tensor
are established for a high temperature dilute gas in two spatial dimensions.
The Chapman-Enskog procedure to first order in the gradients is applied
in order to obtain the dissipative energy and momentum fluxes from
the relativistic Boltzmann equation. The solution for such equation
is written in terms of three sets of orthogonal
polynomials which are explicitly obtained for this calculation. As
in the three dimensional scenario, the heat flux is shown to be
driven by the density, or pressure, gradient additionally to the usual
temperature gradient given by Fourier's law. For the stress (Navier)
tensor one finds, also in accordance with the three dimensional case,
a non-vanishing bulk viscosity for the ideal monoatomic relativistic
two-dimensional gas. All transport coefficients are calculated analytically
for the case of a hard disk gas and the non-relativistic limit for
the constitutive equations is verified.
\end{abstract}


\maketitle

\section{Introduction}\label{s1}

The study of relativistic gases, both from the point of view of their
hydrodynamic and transport properties as well as in relation to the
corresponding thermodynamic and kinetic foundations, has been around
for decades. Moreover, the first works concerning the equilibrium
states of high temperature gases as well as the discussion over the
nature of temperature in relativistic systems date back to over a
century ago \cite{natureFarias}. However, the interest in this kind
of systems has seen a substantial increase in recent years. This is
mostly due to its many applications, some of them arising in present
years due to new theoretical, observational and experimental developments.
Relativistic systems are relevant in astrophysics for example in gamma
ray bursts and some cosmological frameworks. In laboratory physics,
relativistic phenomena are of interest in high temperature plasmas,
heavy ion colliders, lasers and even in graphene, to name a few. Among
these applications, 2D models are relevant in all cases,
for example in accretion disks and other astrophysical systems with
axial symmetry, which are common due to the prevalence of magnetic
fields in the Universe. Moreover, it has been recently pointed out that
relativistic kinetic theory in two dimensions might prove to be useful
in the modeling and analysis of electron flows in two-dimensional materials,
like graphene \cite{Mendoza2013,mendozapress}. Additionally, analytical models of complex
systems in two spatial dimensions are of value since they serve as
a link between theoretical predictions and numerical results in account
of simulations being more easily carried out in lower dimensional
scenarios. 

High temperature dilute gases are considered relativistic fluids for non-negligible
values of their relativistic parameter $z=kT/mc^{2}$, meaning that
the characteristic thermal energy for the system is at least comparable
to the rest energy of its constituents. In such a situation, the relativistic
corrections in the dynamics of the individual particles translate,
upon statistical averages, to measurable modifications to the transport
properties of the system as a whole. Indeed, relativistic kinetic
theory leads to transport equations which can be shown to feature
new, purely relativistic terms impacting both dissipation and inertia.
Additionally, the constitutive equations for the dissipative fluxes
are modified, not only by presenting corrected values of the transport
coefficients but also by containing new terms. In particular, the
heat flux can be driven by density gradients in the system, and the
bulk viscosity term does not vanish for an ideal monoatomic gas as
is the case for mild temperatures.

Early developments of relativistic kinetic theory, including the establishment
of transport coefficients, can be found for example in Refs. \cite{israel63,degroot71},
while Refs.\cite{DeGroot,CercignaniKremer} provide a complete and
clear account of the advances and results in relativistic kinetic
theory up to almost the turn of the century. More recent calculations
introduce the concept of chaotic velocity and address the so-called
generic instability problem, which somehow caused the first order
in the gradients theories to be partially abandoned for decades (see
for example Ref. \cite{JNNFM2010} and references cited therein). The
particular case of two dimensional systems has only recently been
addressed within this framework; in particular M. Miller et. al. \cite{millerUR} establish
thermal and viscous dissipation parameters in the ultrarelativistic
limit using a relaxation time approximation, both in Landau and Eckart's
frame. However, the analytical expressions for the
corresponding coefficients obtained from Boltzmann's equation using the Chapman-Enskog procedure
for the complete range of values of $z$ cannot be currently found in the literature \citep{mendozapress},
other than recent work which only addresses the thermal component of dissipation which
can be found in Ref. \cite{HFBidi}. Moreover, some numerical experiments favor the Chapman-Enskog method as being the most suitable for treating relativistic dissipation within this framework \cite{mendozaCE}.

The purpose of the present work is to obtain the complete set of transport
coefficients for a dilute two-dimensional single component gas, as a function
of the relativistic parameter, from the complete Boltzmann equation. The explicit
values for a hard disks' interaction model are also obtained and benchmarked
against the non-relativistic values in the low temperature limit.
The procedure here followed is based on the Chapman-Enskog expansion
to first order in the Knudsen parameter, which corresponds to the
Navier-Stokes relativistic regime, in a Minkowski 2+1 space-time.
The results regarding thermal dissipation have already been established
in Ref. \cite{HFBidi} for the 2D gas and are here included for the
sake of completeness.

To accomplish such task, the rest of the work is organized as follows.
Section II is devoted to the general setup of the problem where the
space-time metric and phase space variables are specified. Also, the
Boltzmann equation is stated together with the definition of the state
variables and corresponding balance equations according to the Chapman-Enskog
hypothesis. The standard treatment is carried out in section III in
order to recast the kinetic equation in four separate integral equations
subject to subsidiary conditions. The constitutive relations are
addressed in section \ref{s4} IV where polynomial expansions for the solutions
of the integral equations are proposed and transport coefficients
are written in terms of the corresponding parameters. Sections V
and VI are devoted to the explicit calculation of the bulk and shear
viscosities respectively, including the final expressions for the
particular case of a hard disks model. Section VI includes an outline
of the procedure and results carried out in Ref. \cite{HFBidi} for
the coefficients related to heat dissipation. A discussion of the
results and some final remarks are included in section VIII while
the relevant details of some mathematical steps can be found in appendices
A-D. 

\section{Boltzmann equation and Chapman-Enskog approximation}\label{s2}

The physical system to be addressed in the present work, as specified in the previous
section, corresponds to a two-dimensional dilute gas of structureless
particles of mass $m$. The flat Minkowski space-time adequate for
the description for such a system is given by the metric 
\begin{equation}
ds^{2}=c^{2}dt^{2}-dx^{2}-dy^{2},\label{1}
\end{equation}
for which the position and velocity tensors are given by 
\begin{equation}
x^{\nu}=\left(\begin{array}{c}
ct\\
x\\
y
\end{array}\right),\qquad v^{\nu}=\frac{dx^{\nu}}{d\tau}=\gamma_{w}\left(\begin{array}{c}
c\\
w_{x}\\
w_{y}
\end{array}\right),\label{2}
\end{equation}
where $\gamma_{w}=\left(1-\frac{w^{2}}{c^{2}}\right)^{-1/2}$ , with
$w^{2}=-w^{\ell}w_{\ell}$, is the usual Lorentz factor and $\tau$
the particle's proper time. Here and in the rest of this work latin
indices run over $\left\{ 1,2\right\} $ while greek ones do so over
$\left\{ 0,1,2\right\} $ and Einstein's summation convention is implied over repeated indices. 
In this framework, one can define a single particle distribution
function $f\left(x^{\nu},v^{\nu}\right)$, such that
\[
f\left(x^{\nu},v^{\nu}\right)d^{2}xdv^{*},
\]
represents the number of particles at time $t$ occupying a cell of phase space
with volume
$d^{2}xdv^{*}$, where $dv^{*}=d^{2}v/v^{0}$ is the invariant velocity element \cite{DeGroot,CercignaniKremer}. 
The corresponding
Boltzmann equation is given by
\begin{equation}
v^{\alpha}\frac{\partial f}{\partial x^{\alpha}}=\int\int\left(\tilde{f}\tilde{f_{1}}-ff_{1}\right)F
\sigma d\chi dv_{1}^{*},\label{3}
\end{equation}
where the integral operator on the right hand side of Eq.
(\ref{3}) corresponds to the collision kernel. Here a
tilde indicates the corresponding quantity after a collision, $f_1=f\left(x^{\nu},v_{1}^{\nu}\right)$, $F$ denotes the two dimensional
invariant flux, $\sigma$ is the impact parameter and $d\chi$ the
scattering angle element. The corresponding equilibrium distribution
function for the system is a J\"uttner distribution in two dimensions, which
can be written as

\begin{equation}
f^{\left(0\right)}=\frac{ne^{\frac{1}{z}}}{2\pi c^{2}z\left(1+z\right)}e^{-\frac{u^{\alpha}v_{\alpha}}{zc^{2}}},\label{4}
\end{equation}
and satisfies
\begin{equation}
n=\frac{u^{\alpha}}{c^{2}}\int f^{\left(0\right)}v_{\alpha}dv^{*}.\label{5}
\end{equation}
Also, the usual definitions for the hydrodynamic velocity and internal
energy per unit mass hold, namely
\begin{equation}
nu^{\alpha}=\int f^{\left(0\right)}v^{\alpha}dv^{*},\label{6}
\end{equation}
\begin{equation}
n\epsilon= \frac{m u^{\alpha}u^{\beta}}{c^2}\int f^{\left(0\right)}v_{\alpha}v_{\beta}dv^{*}.\label{7}
\end{equation}
For this 2D case one obtains 
\begin{equation}
n\epsilon=nmc^{2}zg\left(z\right)\qquad\textrm{with} \qquad g\left(z\right)=\frac{2z^{2}+2z+1}{z\left(z+1\right)},\label{8}
\end{equation}
and $g\left(z\right)\sim z^{-1}+1$ in the low temperature
limit, yielding the total internal energy (including the rest energy
contribution) for the system. The corresponding definition for the
local equilibrium temperature is given by the relation in Eq.
(\ref{8}). In Eqs. (\ref{4}-\ref{7}), as well as in most of the rest of this
work, the dependence of state variables $n$, $u^{\nu}$, $\epsilon$
and $T$, with space-time, as well as of the distribution function with $v^\mu$ (explicit) and 
with $x^\mu$ (only trough the state variables) is omitted in order to simplify the notation.

Following the Chapman-Enskog procedure to obtain
successive approximations to the solution of the Boltzmann equation,
one assumes the distribution function can be expressed as a series
expansion, with  Knudsen's number as the order parameter, namely
the ratio of the length scale of the gradients in the system
to the characteristic size of the system itself \cite{Ch-E} . Thus, to first order in
such parameter, corresponding to the Navier-Stokes regime, one assumes
\begin{equation}
f=f^{\left(0\right)}\left(1+\phi\left(v^{\nu}\right)\right).
\label{9}
\end{equation}
Introducing this  hypothesis in Eq. (\ref{3}) and linearizing the
collisional term on the right hand side, one obtains an integral equation
for $\phi\left(v^{\nu}\right)$, namely

\begin{equation}
f^{\left(0\right)}\mathcal{C}\left(\phi\left(v^{\nu}\right)\right)=v^{\alpha}\frac{\partial 
f^{\left(0\right)}}{\partial x^{\alpha}},\label{10}
\end{equation}
where
\begin{equation}
\mathcal{C}\left(\phi\left(v^{\nu}\right)\right)=\int\int f^{\left(0\right)}\left(v_{1}\right)
\left(\tilde{\phi}_{1}+\tilde{\phi}-\phi_{1}-\phi\right)F\sigma d\chi dv_{1}^{*},\label{11}
\end{equation}
is the linearized collision kernel. Notice that $\mathcal{C}\left(\phi\left(v^{\nu}\right)\right)$
vanishes if $\phi\left(v^{\nu}\right)$ is a collisional invariant.
In particular, by multiplying Eq. (\ref{10}) by $\phi\left(v^{\nu}\right)=m$ or $\phi\left(v^{\nu}\right)=mv^{\nu}$
and integrating over velocity space ($dv^{*}$) one
obtains the conservation equation for the particle density and the
balance equation for the energy momentum tensor, namely
\begin{equation}
N_{,\nu}^{\nu}=0\qquad  T_{,\nu}^{\mu\nu}=0,\label{12}
\end{equation}
with
\begin{equation}
N^{\nu}=\int f v^{\nu}dv^{*},\label{13}
\end{equation}
\begin{equation}
T^{\mu\nu}=m\int f v^{\mu}v^{\nu}dv^{*},\label{14}
\end{equation}
and a comma indicating a partial derivative. By introducing Eq. (\ref{9}) in Eqs. (\ref{13}) and (\ref{14})
one obtains explicit expressions for the conserved tensors in terms
of the state variables and dissipative fluxes namely
\begin{equation}
N^{\nu}=nu^{\nu},\label{15}
\end{equation}
and
\begin{equation}
T^{\mu\nu}=\frac{n\epsilon}{c^{2}}u^{\mu}u^{\nu}+ph^{\mu\nu}+\pi^{\mu\nu}+\frac{1}{c^{2}}q^{\mu}u^{\nu}+
\frac{1}{c^{2}}u^{\mu}q^{\nu},\label{16}
\end{equation}
where $p$ is the hydrostatic pressure, 
\begin{equation}
q^{\mu}=mh^{\mu\nu}u^{\alpha}\int f^{\left(0\right)}\left(v\right)\phi\left(v^{\nu}\right)v_{\alpha}v_{\nu}d^{*}v,\label{17}
\end{equation}
is the heat flux and
\begin{equation}
\pi^{\mu\nu}=mh^{\mu\alpha}h^{\nu\beta}\int f^{\left(0\right)}\left(v\right)\phi\left(v^{\nu}\right)v_{\alpha}v_{\beta}d^{*}v,\label{18}
\end{equation}
is the Navier tensor. Linear irreversible thermodynamics predicts
linear couplings of these fluxes with the thermodynamic forces of
the same tensor rank \cite{Ch-E}. Indeed, for the case of three
spatial dimensions, the coupling of the heat flux with the temperature
and density (or pressure) gradients as well as the coupling of $\pi^{\mu\nu}$
with the velocity gradient's symmetric traceless part and with its
trace (even for the ideal monoatomic gas) were established since the
early 60's (see for example \cite{DeGroot,israel63}). Moreover, we
obtained the corresponding constitutive equation for the heat flux
in the two-dimensional case in Ref. \cite{HFBidi}, consistent with previous
numerical results \cite{Ghodrat1}. In such a work, only the vector
driving forces were considered in the solution to the Boltzmann equation.
In the following section, the standard Chapman-Enskog method is used
to obtain the complete first order in the gradients deviation for
the equilibrium distribution in the 2+1 case, from which both constitutive equations can be established within such approximation.

\section{The integral equations}\label{s3}

In order to express the right hand side of Eq. (\ref{10}) in terms
of the driving forces, namely the spatial gradients of the state variables,
one uses the fact that the equilibrium distribution function depends
on space-time only through the state variables, that is
\begin{equation}
v^{\alpha}f_{,\alpha}^{\left(0\right)}=\left(v^{\beta}h_{\beta}^{\alpha}+\left(\frac{v^{\beta}u_{\beta}}{c^{2}}\right)u^{\alpha}\right)\left(\frac{\partial f^{\left(0\right)}}{\partial n}n_{,\alpha}+\frac{\partial f^{\left(0\right)}}{\partial u^{\mu}}u_{,\alpha}^{\mu}+\frac{\partial f^{\left(0\right)}}{\partial T}T_{,\alpha}\right),\label{19}
\end{equation}
where the molecular velocity tensor has been written
in its irreducible form in the 2+1 representation as \cite{Eckart,negro}
\begin{equation}
v^{\alpha}=v^{\beta}h_{\beta}^{\alpha}+\left(\frac{v^{\beta}u_{\beta}}{c^{2}}\right)u^{\alpha},\label{20}
\end{equation}
where the projector $h^{\alpha\beta}=\eta^{\alpha\beta}-u^{\alpha}u^{\beta}/c^{2}$
has been introduced. The first and second terms in Eq. (\ref{20}) correspond
to the components of the velocity in the direction orthogonal and
parallel to the hydrodynamic velocity respectively.
Moreover, in a each volume element's comoving frame, where $u^{\alpha}=\left[c,\vec{0}\right]$ , one has that $v^{\beta}h_{\beta}^{\alpha}=\left[0,v^{1},v^{2}\right]$
and $\left(v^{\beta}u_{\beta}/c^{2}\right)u^{\alpha}=\left[v^{0},0,0\right]$.
In such local frames, the first term of the decomposition in Eq. (\ref{20}) corresponds
to the spatial components and the second to the temporal one. This corresponds to the so-called 2+1 decomposition. Notice
that, if we define the molecular velocity measured in the comoving
frame defined above as $K^{\mu}=\gamma_{k}\left[c,\vec{k}\right]$, the coefficient
$v^{\beta}u_{\beta}$, which is an invariant, can be written as $\gamma_{k}c^{2}$
and corresponds to the proper energy (per unit mass) of the molecules \cite{Moratto-Perciante}.

As required by the Chapman-Enskog method, the total time derivatives
which appear in Eq. (\ref{19}), are written in terms of the spatial
gradients by introducing the previous order (Euler) conservation equations.
In this case, the Euler regime equations can be written as
\begin{equation}
u^{\alpha}n_{,\alpha}=-nu_{,\alpha}^{\alpha},\label{21}
\end{equation}
\begin{equation}
u^{\alpha}u_{,\alpha}^{\mu}=-\frac{c^{2}}{\left(g\left(z\right)+1\right)}\left(\frac{T_{,\nu}}{T}+
\frac{n_{,\nu}}{n}\right)h^{\mu\nu},\label{22}
\end{equation}
\vspace{.01cm}
\begin{equation}
u^{\alpha}T_{,\alpha}=-k_{p}\left(z\right)Tu_{,\alpha}^{\alpha},\label{23}
\end{equation}
where
\begin{equation}
k_{p}\left(z\right)=\frac{k}{C_{n}}=\frac{\left(z+1\right)}{z}\left(g\left(z\right)+\frac{2}{z+1}\right)^{-1},\label{24}
\end{equation}
with $g\left(z\right)$ given in Eq. (\ref{8}). Thus, by performing the steps described above, the integral equation (\ref{10})
can be written as
\begin{eqnarray}
\mathcal{C}\left(\phi\left(v^{\nu}\right)\right) & =&v^{\beta}h_{\beta}^{\alpha}\left[\left(\dfrac{\gamma_{k}}{z\left(g\left(z\right)+1\right)}-1\right)\left(g\left(z\right)\dfrac{T_{,\alpha}}{T}-\dfrac{n_{,\alpha}}{n}\right)-\dfrac{v^{\mu}}{zc^{2}}u_{\mu,\alpha}\right]\nonumber \\
 & &-\gamma_{k}\left[1+k_{p}\left(z\right)\left(\dfrac{\gamma_{k}}{z}-g\left(z\right)\right)\right]
 u_{,\alpha}^{\alpha}.\label{25}
\end{eqnarray}

In order to propose a solution to Eq. (\ref{25}), one first separates the velocity gradient as follows
\begin{equation}
u_{\mu,\alpha}=\mathring{\sigma}_{\mu\alpha}+w_{\mu\alpha}+\frac{1}{2}h_{\mu\alpha}u_{,\nu}^{\nu},\label{26}
\end{equation}
where
\begin{equation}
\mathring{\sigma}_{\alpha\beta}=h_{\alpha}^{\gamma}h_{\beta}^{\delta}\frac{\left(u_{\gamma,\delta}
+u_{\delta,\gamma}\right)}{2}-\frac{1}{2}h_{\alpha\beta}u_{,\mu}^{\mu},\label{27}
\end{equation}
and
\begin{equation}
w_{\alpha\beta}=h_{\alpha}^{\gamma}h_{\beta}^{\delta}\frac{\left(u_{\gamma,\delta}-u_{\delta,\gamma}\right)}{2},\label{28}
\end{equation}
are the symmetric traceless and antisymmetric parts of $u_{\mu,\alpha}$
respectively. The antisymmetric part vanishes when contracted
with a symmetric tensor and, using also that $v^{\mu}v^{\beta}h_{\beta}^{\alpha}\mathring{\sigma}_{\mu\alpha}=v^{\mu}v^{\alpha}\mathring{\sigma}_{\mu\alpha}$,
one can write
\begin{eqnarray}
C\left(\phi\left(v^{\nu}\right)\right) & =&v^{\beta}h_{\beta}^{\alpha}\left[\left(\dfrac{\gamma_{k}}{z\left(g\left(z\right)+1\right)}-1\right)\left(g\left(z\right)\dfrac{T_{,\alpha}}{T}-\dfrac{n_{,\alpha}}{n}\right)\right]\nonumber \\
 &-&\dfrac{v^{\mu}v^{\alpha}}{zc^{2}}\mathring{\sigma}_{\mu\alpha}+\left[\left(\dfrac{1}{2}-k_{p}\left(z\right)\right)\dfrac{\gamma_{k}^{2}}{z}+\left(k_{p}\left(z\right)g\left(z\right)-1\right)\gamma_{k}-\dfrac{1}{2z}\right]u_{,\nu}^{\nu}.\label{29}
\end{eqnarray}
Taking into account the structure of Eq. (\ref{29}), its general solution is proposed as
follows
\begin{equation}
\phi\left(v^{\nu}\right)=\mathcal{A}_{1}\left(\gamma_{k}\right)v^{\beta}h_{\beta}^{\alpha}\frac{T_{,\alpha}}{T}+\mathcal{A}_{2}\left(\gamma_{k}\right)v^{\beta}h_{\beta}^{\alpha}\frac{n_{,\alpha}}{n}+\mathcal{A}_{3}\left(\gamma_{k}\right)u_{,\alpha}^{\alpha}+\mathcal{A}_{\alpha\beta}\mathring{\sigma}^{\alpha\beta}+\alpha+\check{\alpha}_{\nu}v^{\nu},\label{30}
\end{equation}
where $\alpha$ and $\check{\alpha}_{\nu}$ can only be functions
of the state variables, while the scalar quantities $\mathcal{A}_{i}$
can also depend on $\gamma_{k}$. For uniqueness of the Chapman-Enskog
solution, subsidiary equations need to be supplied, which place restrictions
on the coefficients. Since the local state variables
are defined as averages weighted by the equilibrium distribution function,
the adequate conditions, which guarantee uniqueness of the solution and consistency with the definitions
given by Eqs. (\ref{5})-(\ref{7}) reads:
\begin{equation}
\int f^{(0)}\phi\left(v^{\alpha}\right)\gamma_{k}^{2}dv^{*}=0,\label{31}
\end{equation}
\begin{equation}
\int f^{(0)}\phi\left(v^{\alpha}\right)v^{\mu}dv^{*}=0.\label{32}
\end{equation}
Moreover, by using the properties of the
tensor $\mathring{\sigma}^{\alpha\beta}$ and the integrals in Appendix
A, Eq.(\ref{30}) can be further reduced to (see Appendix B) 
\begin{equation}
\phi\left(v^{\nu}\right)=\left(A_{1}\left(\gamma_{k}\right)\frac{T_{,\alpha}}{T}+A_{2}\left(\gamma_{k}\right)
\frac{n_{,\alpha}}{n}\right)v^{\beta}h_{\beta}^{\alpha}+A_{3}\left(\gamma_{k}\right)u_{,\alpha}^{\alpha}
+A_{4}\left(\gamma_{k}\right)v_{\alpha}v_{\beta}\mathring{\sigma}^{\alpha\beta},\label{33}
\end{equation}
and the subsidiary conditions can be written as
\begin{equation}
\int A_{i}\left(\gamma_{k}\right)\left(1-\gamma_{k}^{2}\right)e^{-\gamma_{k}/z}d\gamma_{k}=0\quad i=1\,2\label{34}
\end{equation}
\begin{equation}
\int A_{3}\left(\gamma_{k}\right)\gamma_{k}e^{-\gamma_{k}/z}d\gamma_{k}=\int A_{3}\left(\gamma_{k}\right)\gamma_{k}^{2}e^{-\gamma_{k}/z}d\gamma_{k}=0,\label{35}
\end{equation}
where we have used that $dv^{*}=2\pi c^{2}d\gamma_{k}$ in the 2D
case. Substituting Eq. (\ref{33}) in Eq. (\ref{25}) and making
use of the fact that in the present representation $n$, $T$ and
$u^{\nu}$ are independent variables, the problem can be separated in
four independent integral equations namely,
\begin{equation}
\mathcal{C}\left(-\frac{A_{1}\left(\gamma_{k}\right)}{g\left(z\right)}v^{\beta}h_{\beta}^{\alpha}\right)=
-\frac{1}{z\left(g\left(z\right)+1\right)}v^{\beta}h_{\beta}^{\alpha}\left(\gamma_{k}-z\left(g\left(z\right)+1\right)\right),\label{36}
\end{equation}

\begin{equation}
\mathcal{C}\left(A_{2}\left(\gamma_{k}\right)v^{\beta}h_{\beta}^{\alpha}\right)=-\frac{1}{z\left(g\left(z\right)+1\right)}v^{\beta}
h_{\beta}^{\alpha}\left(\gamma_{k}-z\left(g\left(z\right)+1\right)\right),\label{37}
\end{equation}

\begin{equation}
\mathcal{C}\left(A_{3}\left(\gamma_{k}\right)\right)=\left(\frac{1}{2}-k_{p}\left(z\right)\right)\frac{\gamma_{k}^{2}}{z}
+\left(k_{p}\left(z\right)g\left(z\right)-1\right)\gamma_{k}-\frac{1}{2z},\label{38}
\end{equation}

\begin{equation}
\mathcal{C}\left(A_{4}\left(\gamma_{k}\right)v^{\alpha}v^{\beta}\left(h_{\alpha}^{\mu}h_{\beta}^{\nu}-
\frac{1}{2}h_{\alpha\beta}h^{\mu\nu}\right)\right)=-\frac{1}{zc^{2}}v^{\alpha}v^{\beta}\left(h_{\alpha}^{\mu}h_{\beta}^{\nu}-
\frac{1}{2}h_{\alpha\beta}h^{\mu\nu}\right),\label{39}
\end{equation}
where we have used the identity
\begin{equation}
\mathring{\sigma}_{\alpha\beta}v^{\alpha}v^{\beta}=v^{\alpha}v^{\beta}\left(h_{\alpha}^{\mu}h_{\beta}^{\nu}-\frac{1}{2}
h_{\alpha\beta}h^{\mu\nu}\right)u_{\mu,\nu},\label{40}
\end{equation}
in order to obtain Eq. (\ref{39}). The corresponding problems will
be addressed in Sects. 5-8. However before proceeding to the solutions
for the unknown functions $A_{i}\left(\gamma_{k}\right)$, it is convenient to analyze the constitutive
relations that arise from Eqs. (\ref{17}) and (\ref{18}). This task will be undertaken in the next section.

\section{The constitutive equations and proposed expansions}\label{s4}

The solution to the integral equations derived in the previous section
are required in order to obtain the transport coefficients relating the dissipative
fluxes (heat and momentum) with each corresponding thermodynamic force,
namely the gradients of the state variables. Indeed, substituting
Eqs. (\ref{9}) and (\ref{33}) in Eqs.(\ref{17})
-(\ref{18}) and invoking Curie's principle, which states that only forces and fluxes of the same tensor
rank can couple in such relations, one obtains
\begin{equation}
q^{\mu}=mh^{\mu\nu}h_{\beta}^{\alpha}c^{2}\int f^{\left(0\right)}\left(A_{1}\left(\gamma_{k}\right)\frac{T_{,\alpha}}{T}
+A_{2}\left(\gamma_{k}\right)\frac{n_{,\alpha}}{n}\right)v^{\beta}\gamma_{k}v_{\nu}d^{*}v,\label{41}
\end{equation}
\begin{eqnarray}
\pi^{\mu\nu} & =&u_{,\alpha}^{\alpha}h^{\mu\nu}\left\{ \frac{m}{2}h_{\alpha\beta}\int f^{\left(0\right)}A_{3}\left(\gamma_{k}\right)v^{\alpha}v^{\beta}d^{*}v\right\} \nonumber \\
 & +&\mathring{\sigma}_{\delta\varepsilon}\left\{ m\left(h_{\alpha}^{\mu}h_{\beta}^{\nu}-\frac{1}{2}h_{\alpha\beta}h^{\mu\nu}\right)
 \int f^{\left(0\right)}A_{4}\left(\gamma_{k}\right)v^{\delta}v^{\varepsilon}v^{\alpha}v^{\beta}d^{*}v\right\}, \label{42}
\end{eqnarray}
where the stress tensor $\pi^{\mu\nu}$ has
been separated into a traceless component and a diagonal (scalar) one
in order to associate to each of them its corresponding driving
term in $f^{\left(1\right)}$. Using the integrals in Appendix A,
one can show that the previous equations can be written as 
\begin{equation}
q^{\mu}=h^{\mu\alpha}\left(L_{T}\frac{T_{,\alpha}}{T}+L_{n}\frac{n_{,\alpha}}{n}\right)\label{43}
\end{equation}
and
\begin{equation}
\pi^{\mu\nu}=\mu h^{\mu\nu}u_{,\alpha}^{\alpha}+\eta\mathring{\sigma}^{\mu\nu},\label{44}
\end{equation}
where the transport coefficients are given by
\begin{equation}
L_{T}=\frac{mnc^{4}e^{1/z}}{2z\left(1+z\right)}\int e^{-\gamma_{k}/z}A_{1}\left(\gamma_{k}\right)
\left(1-\gamma_{k}^{2}\right)\gamma_{k}d\gamma_{k},\label{45}
\end{equation}

\begin{equation}
L_{n}=\frac{mnc^{4}e^{1/z}}{2z\left(1+z\right)}\int e^{-\gamma_{k}/z}A_{2}\left(\gamma_{k}\right)
\left(1-\gamma_{k}^{2}\right)\gamma_{k}d\gamma_{k},\label{46}
\end{equation}

\begin{equation}
\mu=\frac{mnc^{2}e^{1/z}}{2z\left(1+z\right)}\int e^{-\gamma_{k}/z}A_{3}\left(\gamma_{k}\right)
\left(1-\gamma_{k}^{2}\right)d\gamma_{k}=\frac{mne^{1/z}c^{2}}{2z\left(1+z\right)}\int e^{-\gamma_{k}/z}A_{3}
\left(\gamma_{k}\right)d\gamma_{k},\label{47}
\end{equation}

\begin{equation}
\eta=\frac{mne^{1/z}c^{4}}{8z\left(1+z\right)}\int e^{-\gamma_{k}/z}A_{4}\left(\gamma_{k}\right)
\left(\gamma_{k}^{2}-1\right)^{2}d\gamma_{k},\label{48}
\end{equation}
where we have used Eq. (\ref{35}) in order to obtain the second equality in Eq. (\ref{47}).
Here $L_{T}$ corresponds to the relativistic thermal conductivity
while $L_{n}$ is a purely relativistic coefficient which couples
the heat flux with the density gradient in this representation \cite{Micro,JNNFM2010}.
The coefficients in Eqs. (\ref{47}) and (\ref{48}) correspond to
the bulk and shear viscosities respectively. The non-relativistic
values for $L_{T}$ and $\eta$ in the case of the gas of diameter
$d$ hard disks are given by \cite{Sengers}
\[
L_{T\left(\text{NR}\right)}=\frac{2mc^{3}}{\sqrt{\pi}}\frac{z^{3/2}}{d}\qquad\text{and}\qquad
\eta_{\left(\text{NR}\right)}=\frac{mc}{2\sqrt{\pi}}\frac{\sqrt{z}}{d},
\]
respectively, while $L_{n}$ and $\mu$ have no non-relativistic
counterpart.

Following the standard procedure, the solutions to Eqs. (\ref{36})-(\ref{39}) are expressed as series expansions of appropriate orthogonal
polynomials.
In view of the structure of the expressions for the transport coefficients,
the following series are considered:
\begin{equation}
A_{i}\left(\gamma_{k}\right)=\sum_{n=1}^{\infty}a_{n}^{\left(i\right)}\mathcal{P}_{n}^{\left(i\right)}\left(\gamma_{k}\right),\label{49}
\end{equation}
with
\begin{equation}
\mathcal{P}_{n}^{\left(i\right)}\left(\gamma_{k}\right)=\sum_{j=0}^{n}\alpha_{nj}^{\left(i\right)}\gamma_{k}^{j},\label{50}
\end{equation}
and
\begin{equation}
\int\mathcal{P}_{n}^{\left(i\right)}\left(\gamma_{k}\right)\mathcal{P}_{m}^{\left(i\right)}\left(\gamma_{k}\right)p^{\left(i\right)}
\left(\gamma_{k}\right)d\gamma=\delta_{nm},\label{51}
\end{equation}
where $p^{\left(1\right)}\left(\gamma_{k}\right)=p^{\left(2\right)}\left(\gamma_{k}\right)=e^{-\gamma_{k}/z}\left(1-\gamma_{k}^{2}\right)$,
$p^{\left(3\right)}\left(\gamma_{k}\right)=e^{-\gamma_{k}/z}$ and
$p^{\left(4\right)}\left(\gamma_{k}\right)=e^{-\gamma_{k}/z}\left(1-\gamma_{k}^{2}\right)^{2}$.
Introducing Eq. (\ref{49}) in Eqs. (\ref{45})-(\ref{48}) and using
the orthogonality condition given in Eq. (\ref{51}) leads to expressions
for the transport coefficients which depend only on one coefficient
of each series namely, 
\begin{equation}
L_{T}=\frac{mnc^{4}e^{1/z}}{2z\left(1+z\right)}\frac{a_{1}^{\left(1\right)}}{\alpha_{11}^{\left(1\right)}},\label{52}
\end{equation}
\begin{equation}
L_{n}=\frac{mnc^{4}e^{1/z}}{2z\left(1+z\right)}\frac{a_{1}^{\left(2\right)}}{\alpha_{11}^{\left(2\right)}},\label{53}
\end{equation}
\begin{equation}
\mu=\frac{mne^{1/z}c^{2}}{2z\left(1+z\right)}\frac{a_{0}^{\left(3\right)}}{\alpha_{00}^{\left(3\right)}},\label{54}
\end{equation}
\begin{equation}
\eta=\frac{mne^{1/z}c^{4}}{8z\left(1+z\right)}\frac{a_{0}^{\left(4\right)}}{\alpha_{00}^{\left(4\right)}}.\label{55}
\end{equation}
The values for the coefficients $\alpha_{jn}^{\left(i\right)}$
required in the previous equations and in the remainder of this work, have been obtained by carrying out a standard 
Gram-Shmidt procedure, and are here listed in Appendix C. 

As one final step before proceeding to the solution of
the integral equations, it is advantageous to establish the conditions
that Eqs. (\ref{34}) and (\ref{35}) imply over the coefficients
$a_{n}^{\left(i\right)}$. Clearly, Eq. (\ref{34}) leads to $a_{0}^{\left(1\right)}=a_{0}^{\left(2\right)}=0$.
On the other hand, in order to enforce the condition in Eq. (\ref{35})
on the coefficients $a_{n}^{\left(3\right)}$, one writes
\begin{equation}
\gamma_{k}=\sqrt{\frac{z}{e^{1/z}}}\left[z\mathcal{P}_{1}^{\left(3\right)}\left(\gamma_{k}\right)+\left(1+z\right)
\mathcal{P}_{0}^{\left(3\right)}\left(\gamma_{k}\right)\right],\label{56}
\end{equation}
and 
\begin{equation}
\gamma_{k}^{2}=\sqrt{\frac{z}{e^{1/z}}}\left[2z^{2}\mathcal{P}_{2}^{\left(3\right)}\left(\gamma_{k}\right)+2z
\left(1+2z\right)\mathcal{P}_{1}^{\left(3\right)}\left(\gamma_{k}\right)+\left(1+2z+2z^{2}\right)\mathcal{P}_{0}^{\left(3\right)}
\left(\gamma_{k}\right)\right],\label{57}
\end{equation}
which can be shown to lead to the following equations
\begin{equation}
a_{1}^{\left(3\right)}=-\frac{1+z}{z}a_{0}^{\left(3\right)},\label{58}
\end{equation}
\begin{equation}
2z^{2}a_{2}^{\left(3\right)}+2z\left(1+2z\right)a_{1}^{\left(3\right)}+\left(1+2z+2z^{2}\right)a_{0}^{\left(3\right)}=0.\label{59}
\end{equation}
The next three sections are devoted to finding approximate solutions to Eqs. (\ref{36})-(\ref{39}) which, once introduced in 
Eqs.  (\ref{52})-(\ref{55}) yield integral expressions for the transport coefficients.

\section{The scalar equation and the bulk viscosity}\label{s5}

The solution to Eq. (\ref{38}) yields the coefficient required to calculate the bulk viscosity, whose nature is purely relativistic
in the case of the monoatomic ideal gas. In order to obtain a first
approximation to $a_{0}^{\left(3\right)}$ one starts by substituting
Eq. (\ref{49}) in Eq. (\ref{38}) and using Eqs. (\ref{56}) and
(\ref{57}), by means of which the corresponding integral equation
can be written as
\begin{equation}
\sum_{n=0}^{\infty}a_{n}^{\left(3\right)}\mathcal{C}\left(\mathcal{P}_{n}^{\left(3\right)}
\left(\gamma_{k}\right)\right)=\left(1-2k_{p}\left(z\right)\right)\sqrt{\frac{z}{e^{1/z}}}z
\mathcal{P}_{2}^{\left(3\right)}\left(\gamma_{k}\right).\label{60}
\end{equation}
Multiplying both sides of Eq. (\ref{60}) by $\mathcal{P}_{m}^{\left(3\right)}f^{\left(0\right)}$
and integrating over velocity space yields
\begin{equation}
-\sum_{n=0}^{\infty}\sum_{\ell=0}^{n}\sum_{j=0}^{2}\alpha_{n\ell}^{\left(3\right)}\alpha_{2j}^{\left(3\right)}
a_{n}^{\left(3\right)}\left[\gamma_{k}^{\ell},\gamma_{k}^{j}\right]=\frac{e^{1/z}}{n\left(1+z\right)}\left(1-2k_{p}
\left(z\right)\right)\sqrt{\frac{z}{e^{1/z}}},\label{61}
\end{equation}
where the collisional bracket is defined in the usual fashion:
\begin{equation}
\left[H,G\right]=-\frac{1}{n^{2}}\int\mathcal{C}\left(H\right)Gf^{\left(0\right)}d^{*}v,\label{62}
\end{equation}
and satisfies
\begin{equation}
\left[H,G\right]=-\frac{1}{4n^{2}}\int\left(H'+H'_{1}-H-H_{1}\right)\left(G'+G'_{1}-G-G_{1}\right)f^{(0)}f_{*}^{(0)}F
\sigma(\chi)d\chi dv_{1}^{*}dv^{*}d^{*}v.\label{63}
\end{equation}
Notice that this identity leads to $\left[\gamma_{k}^{i},\gamma_{k}^{j}\right]=0$ if
$i,j=0,1$ due to the conservation properties of elastic binary collisions.
The first approximation to $a_{0}^{\left(3\right)}$ arises from considering the first term of
the sum on the left hand side of Eq. (\ref{61}), in this case being
$n=2$, which yields
\begin{equation}
a_{2}^{\left(3\right)}=-\left(\alpha_{22}^{\left(3\right)}\right)^{-2}\frac{\sqrt{ze^{1/z}}}{n\left(1+z\right)}\left(1-2k_{p}
\left(z\right)\right)\left[\gamma_{k}^{2},\gamma_{k}^{2}\right]^{-1}.\label{64}
\end{equation}
Notice that the coefficient required for the coefficient $\mu$ is
$a_{0}^{\left(3\right)}$ while the first approximation leads to $a_{2}^{\left(3\right)}$.
However, Eqs. (\ref{58}) and (\ref{59}) allows one to solve for $a_{0}^{\left(3\right)}$
as follows
\begin{equation}
a_{0}^{\left(3\right)}=\frac{2z^{2}a_{2}^{\left(3\right)}}{1+4z+2z^{2}},\label{65}
\end{equation}
and thus, by introducing the expressions for $\alpha_{ii}^{\left(3\right)}$
and considering $\left[\gamma_{k}^{2},\gamma_{k}^{2}\right]$ for a hard disks gas (see Appendices
C and D respectively), one obtains from Eq. (\ref{54}),
\begin{equation}
\mu=\frac{30mc}{d}\frac{z^{6}}{\left(1+4z+2z^{2}\right)^{2}}\mathcal{I}_{1}^{-1},\label{66}
\end{equation}
with
\begin{equation}
\mathcal{I}_{1}=\int_{2/z}^{\infty}e^{-\left(x-\frac{2}{z}\right)}\left(\frac{1}{x}+\frac{3}{x^{2}}+\frac{3}{x^{3}}\right)
\left(z^{2}x^{2}-4\right)^{5/2}dx,\label{67}
\end{equation}
(see Figures \ref{fig:1} and \ref{fig:2}).

\section{The tensor equation and the shear viscosity}\label{s6}
In order to obtain the coefficient $a_{0}^{\left(4\right)}$, we start
by writing Eq. (\ref{39}) as
\begin{equation}
\sum_{n=0}^{\infty}a_{n}^{\left(4\right)}C\left(\mathcal{P}_{n}^{\left(4\right)}\left(\gamma_{k}\right)\mathring{v^{\mu}v^{\nu}}
\right)=-\frac{1}{zc^{2}\alpha_{00}^{\left(4\right)}}\mathring{v^{\mu}v^{\nu}}\mathcal{P}_{0}^{\left(4\right)}
\left(\gamma_{k}\right).\label{68}
\end{equation}
where 
\begin{equation}
\mathring{v^{\mu}v^{\nu}}=v^{\alpha}v^{\beta}\left(h_{\alpha}^{\mu}h_{\beta}^{\nu}-\frac{1}{2}h_{\alpha\beta}h^{\mu\nu}\right),\label{69}
\end{equation}
is the traceless velocity dyad and we have used Eq. (\ref{49}). Multiplying
both sides by $\mathcal{P}_{m}^{\left(4\right)}\left(\gamma\right)\mathring{v_{\mu}v_{\nu}}f^{(0)}$
and integrating yields
\begin{equation}
\sum_{n=0}^{\infty}\sum_{i=0}^{n}a_{n}^{(4)}\alpha_{ni}^{\left(4\right)}\left[\gamma^{i}\mathring{v^{\mu}v^{\nu}},
\gamma^{j}\mathring{v_{\mu}v_{\nu}}\right]=\frac{1}{2}\frac{ne^{1/z}}{z^{2}\left(1+z\right)}
\frac{c^{2}}{\left(\alpha_{00}^{\left(4\right)}\right)^{2}},\label{70}
\end{equation}
where we have used the identity
\begin{equation}
\mathring{v^{\mu}v^{\nu}}\mathring{v_{\mu}v_{\nu}}=v^{\alpha}v^{\beta}v^{\lambda}v^{\sigma}\left(h_{\lambda\alpha}h_{\sigma\beta}
-\frac{1}{2}h_{\lambda\sigma}h_{\alpha\beta}\right)=\frac{1}{2}\left(1-\gamma^{2}\right)^{2}c^{4}.\label{71}
\end{equation}
Using once again the symmetries of the linearized collisional kernel,
one obtains the following expression for the first approximation to
$a_{0}^{\left(4\right)}$
\begin{equation}
a_{0}^{(4)}=\frac{1}{n}\frac{e^{1/z}}{z^{2}\left(1+z\right)}\frac{c^{2}}{\left(\alpha_{00}^{\left(4\right)}\right)^{3}}
\left(2\left[v^{\mu}v^{\nu},v_{\mu}v_{\nu}\right]-4c^{2}\left[\gamma v^{\mu},\gamma v_{\mu}\right]+c^{4}\left[\gamma^{2},
\gamma^{2}\right]\right)^{-1}.\label{72}
\end{equation}
Substitution of the values for the brackets values obtained in Appendix
D one obtains, for the shear viscosity
\begin{equation}
\eta=\frac{60mc}{d}z^{2}\left(3z^{2}+3z+1\right)^{2}\mathcal{I}_{2}^{-1},\label{73}
\end{equation}
where
\begin{equation}
\mathcal{I}_{2}=\int_{2/z}^{\infty}e^{-\left(x-\frac{2}{z}\right)}
\left(2+x+\frac{3}{x}+\frac{3}{x^{2}}+\frac{3}{x^{3}}\right)\left(z^{2}x^{2}-4\right)^{5/2}dx,\label{74}
\end{equation}
(see Figure \ref{fig:1}).

\section{The vector equation and the heat flux}
In this section, only the main steps of the procedure to obtain the
transport coefficients corresponding to the heat flux constitutive
equation are shown. The calculation follows somewhat closely the one
carried out in Ref. \cite{HFBidi} and the reader is referred to such
publication for further details. By direct inspection one notices
that the integral equations (\ref{35}) and (\ref{37}) have the same
structure and thus only the first one needs to be solved. The solution
for $A_{2}\left(\gamma_{k}\right)$ will then be obtained simply as
$-A_{1}\left(\gamma_{k}\right)/g\left(z\right)$. We thus start by
multiplying Eq. (\ref{36}) by $v^{\nu}h_{\nu\alpha}$ and inserting
the proposed solution given by Eq. (\ref{49}), which yields
\begin{equation}
\sum_{n=1}^{\infty}a_{n}^{\left(1\right)}\mathcal{C}\left(\mathcal{P}_{n}^{\left(1\right)}\left(\gamma_{k}\right)v^{\beta}
h_{\beta}^{\alpha}\right)v^{\nu}h_{\nu\alpha}=\frac{c^{2}}{\alpha_{11}^{\left(1\right)}}\frac{g\left(z\right)}{z\left(g\left(z\right)
+1\right)}\mathcal{P}_{1}^{\left(1\right)}\left(\gamma_{k}\right)\left(1-\gamma_{k}^{2}\right),\label{75}
\end{equation}
where we have used $a_{0}^{(1)}=0$, in view of Eq. (\ref{33}). In
order to obtain the first approximation for $a_{1}^{\left(1\right)}$,
one further multiplies both sides by $\mathcal{P}_{m}^{(1)}\left(\gamma_{k}\right)f^{\left(0\right)}$,
which upon integration over $dk^{*}$ yields
\begin{equation}
a_{1}^{(1)}=-\frac{e^{\frac{1}{z}}\left(\alpha_{11}^{\left(1\right)}\right)^{-3}c^{2}g\left(z\right)}{nz^{2}\left(1+z\right)
\left(g\left(z\right)+1\right)}\left(\left[\gamma_{k}v^{\alpha},\gamma_{k}v_{\alpha}\right]-c^{2}\left[\gamma_{k}^{2},
\gamma_{k}^{2}\right]\right)^{-1}.\label{76}
\end{equation}
The collision brackets in Eq. (\ref{76}) are calculated in Appendix
C. Introducing such results yields
\begin{equation}
a_{1}^{(1)}=\frac{15}{d}\frac{\left(\alpha_{1}^{\left(1\right)}\right)^{-3}g\left(z\right)\left(1+z\right)e^{\frac{1}{z}}}{n
\left(g\left(z\right)+1\right)cz^{3}}\mathcal{I}_{3}^{-1},\label{77}
\end{equation}
with
\begin{equation}
\mathcal{I}_{3}=\int_{2/z}^{\infty}e^{-\left(x-\frac{2}{z}\right)}\left\{ 1+\frac{3}{x}+\frac{6}{x^{2}}+\frac{6}{x^{3}}\right\} 
\left(z^{2}x^{2}-4\right)^{5/2}dx,\label{78}
\end{equation}
which leads to the following expressions for the transport coefficients:
\begin{equation}
L_{T}=\frac{30mc^{3}}{d}\frac{z^{4}\left(3z^{2}+6z+2\right)^{2}}{\left(1+z\right)^{2}}\frac{\left(2z^{2}+2z+1\right)}{
\left(3z^{2}+3z+1\right)}\mathcal{I}_{3}^{-1},\label{79}
\end{equation}
\begin{equation}
L_{nT}=-\frac{30mc^{3}}{d}\frac{z^{5}\left(3z^{2}+6z+2\right)^{2}}{\left(3z^{2}+3z+1\right)\left(1+z\right)}\mathcal{I}_{3}^{-1},\label{80}
\end{equation}
which are the expressions obtained in Ref. \cite{HFBidi} (see Figures \ref{fig:1}-\ref{fig:3}).

\section{Discussion and final remarks}

\begin{figure}
\includegraphics[scale=0.6]{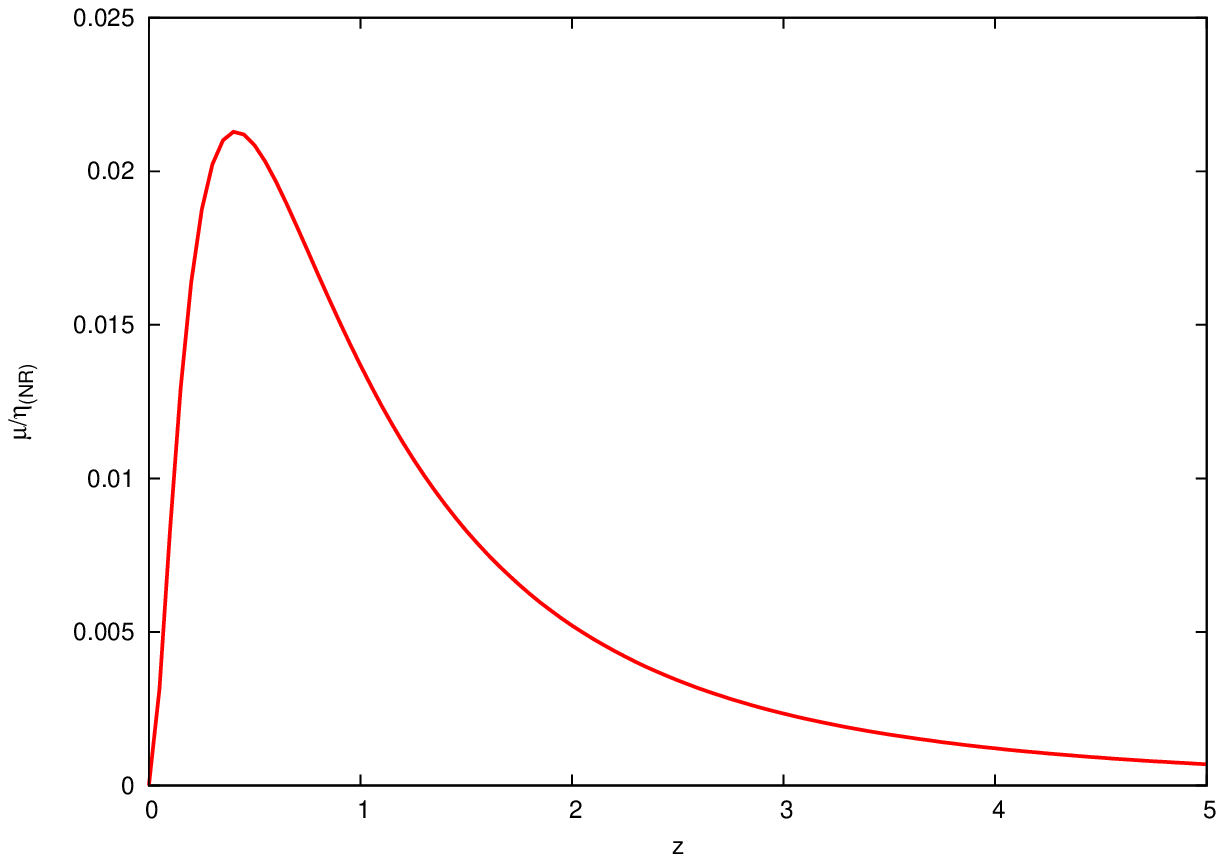}\includegraphics[scale=0.6]{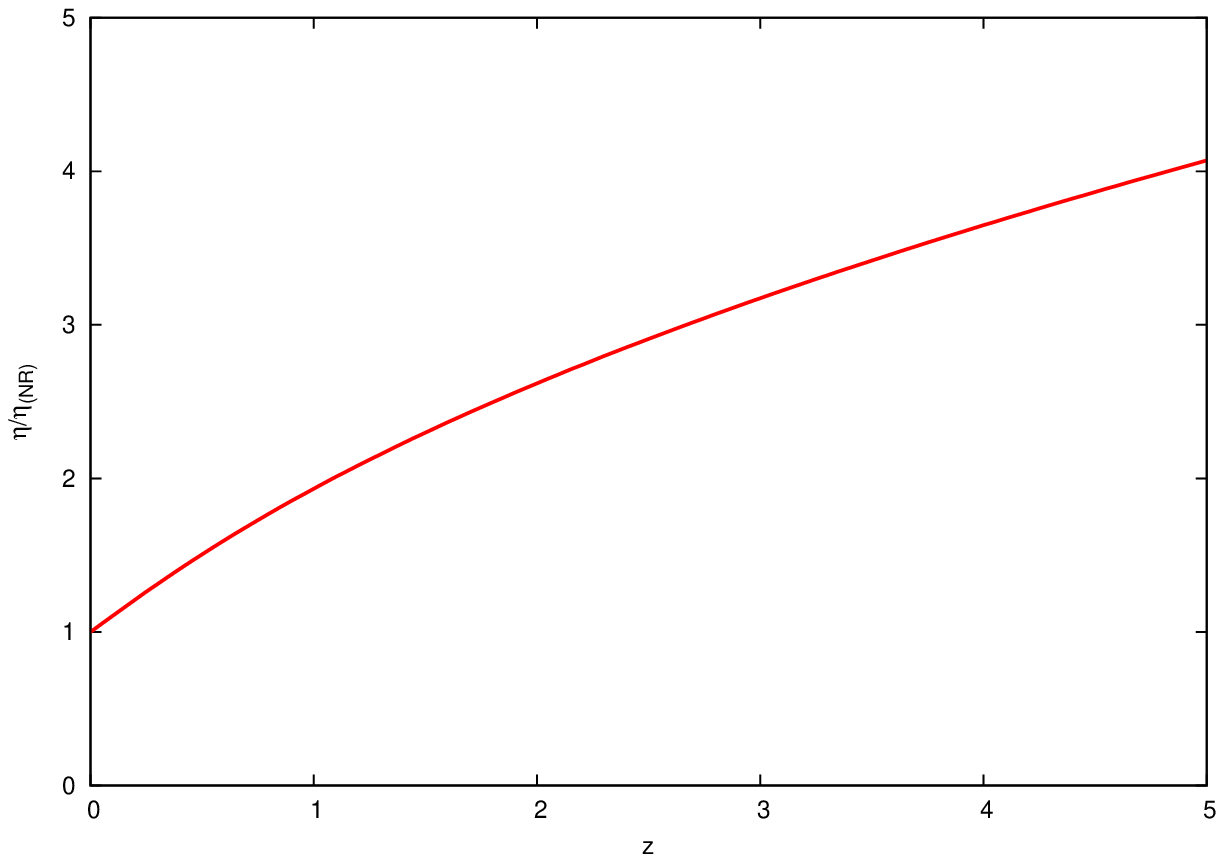}\\
\includegraphics[scale=0.6]{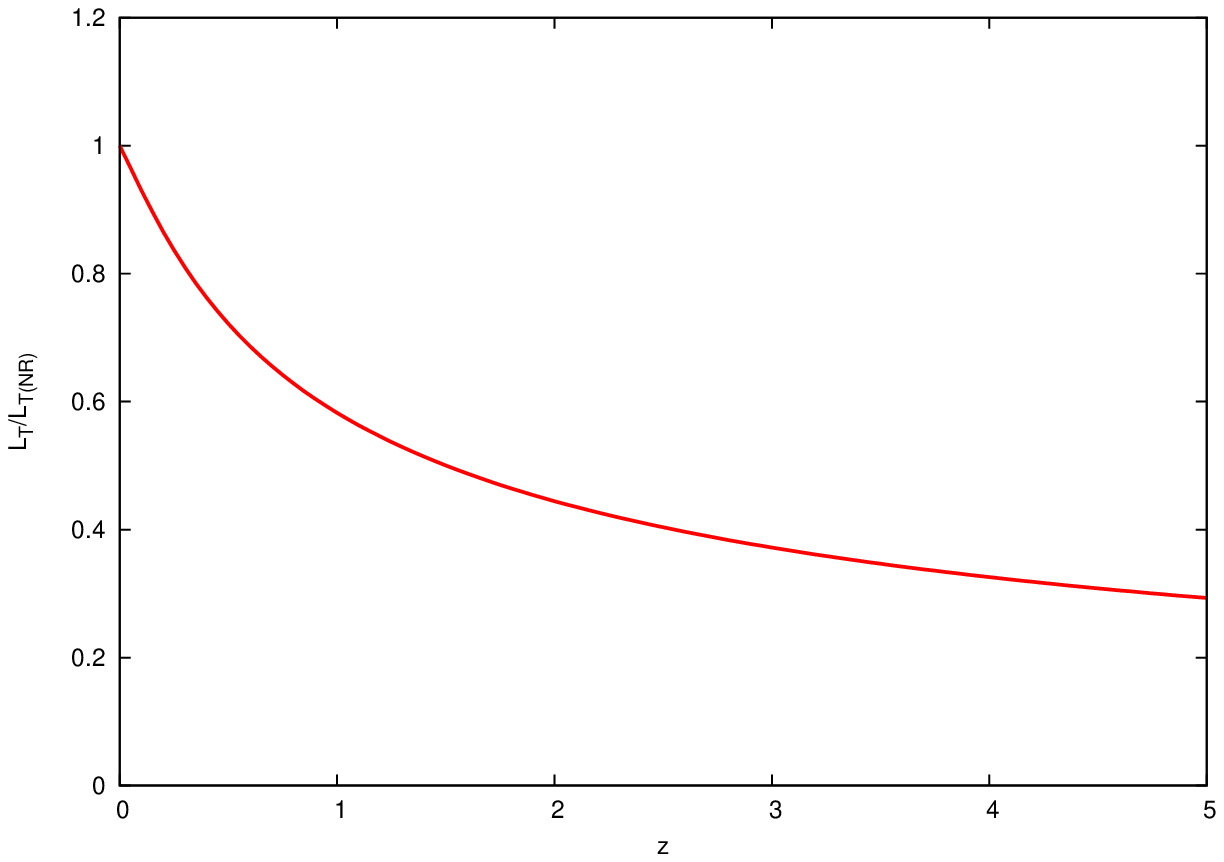}\includegraphics[scale=0.6]{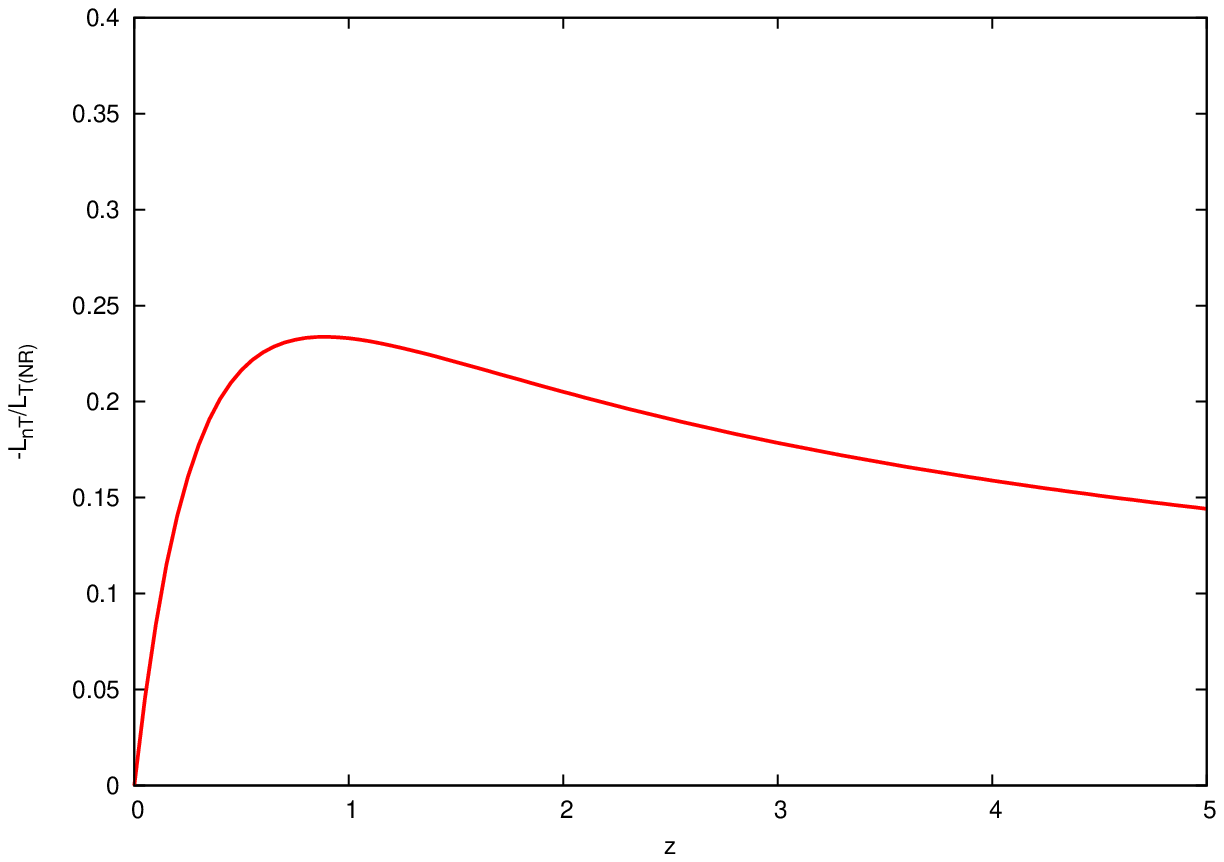}
\caption{\label{fig:1}The transport coefficients $\mu/\eta_{\left(NR\right)}$,
$\eta/\eta_{\left(NR\right)}$, $L_{T}/L_{T(NR)}$ and $L_{nT}/L_{T(NR)}$
from left to right - up to down, respectively.}
\end{figure}

In the previous sections, the Chapman-Enskog method was used in order
to obtain first order approximation to the transport coefficients
of a two-dimensional relativistic gas from the complete Boltzmann equation.
The analytical expressions obtained in terms of collision brackets
are general and can be evaluated for particular models of molecular
interactions. Explicit values for the thermal and viscous coefficients
were established in Sects. 5-7 for the particular case of a hard disks
system. The functional behavior of these are plotted in Fig. (\ref{fig:1}),
where $L_{T}$ and $L_{nT}$ are normalized to the non-relativistic
thermal conductivity and $\mu$ and $\eta$ to the non-relativistic
shear viscosity.

\begin{figure}
\includegraphics[scale=0.6]{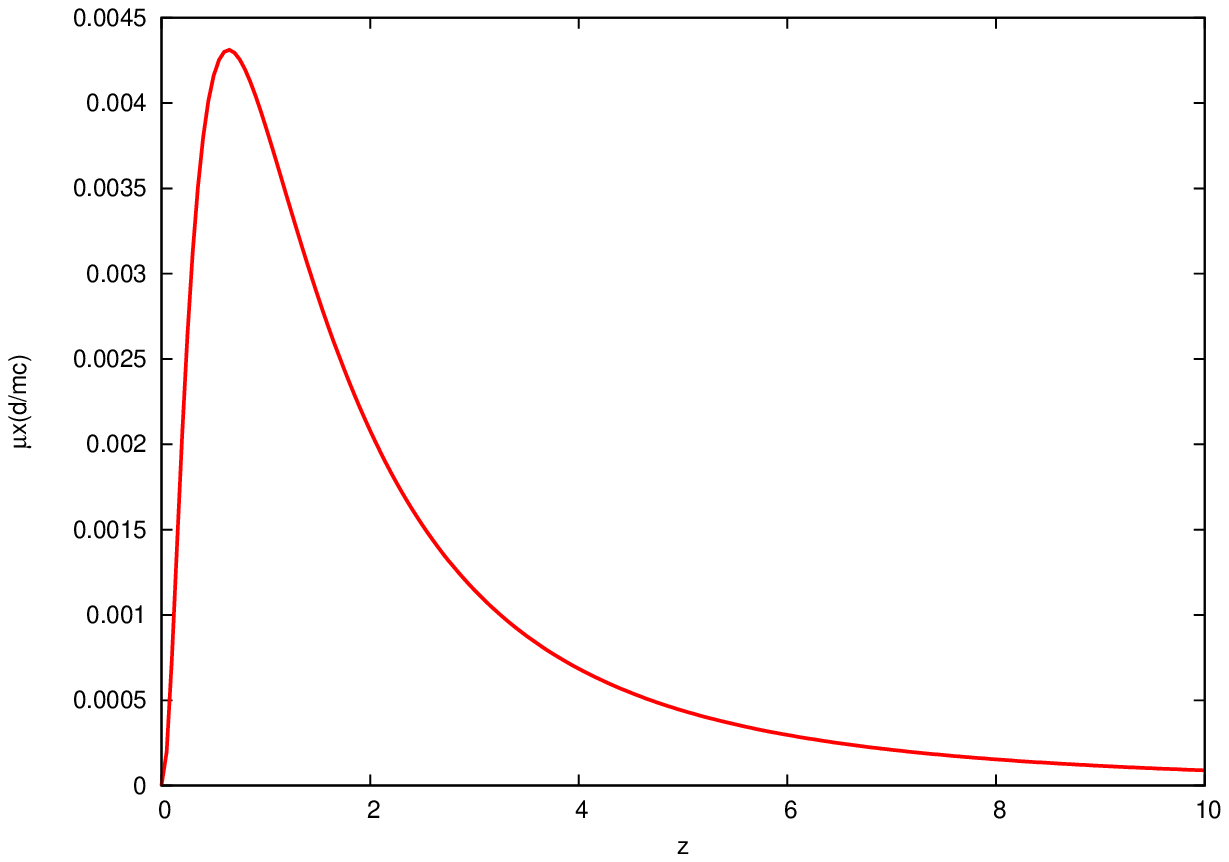}\includegraphics[scale=0.6]{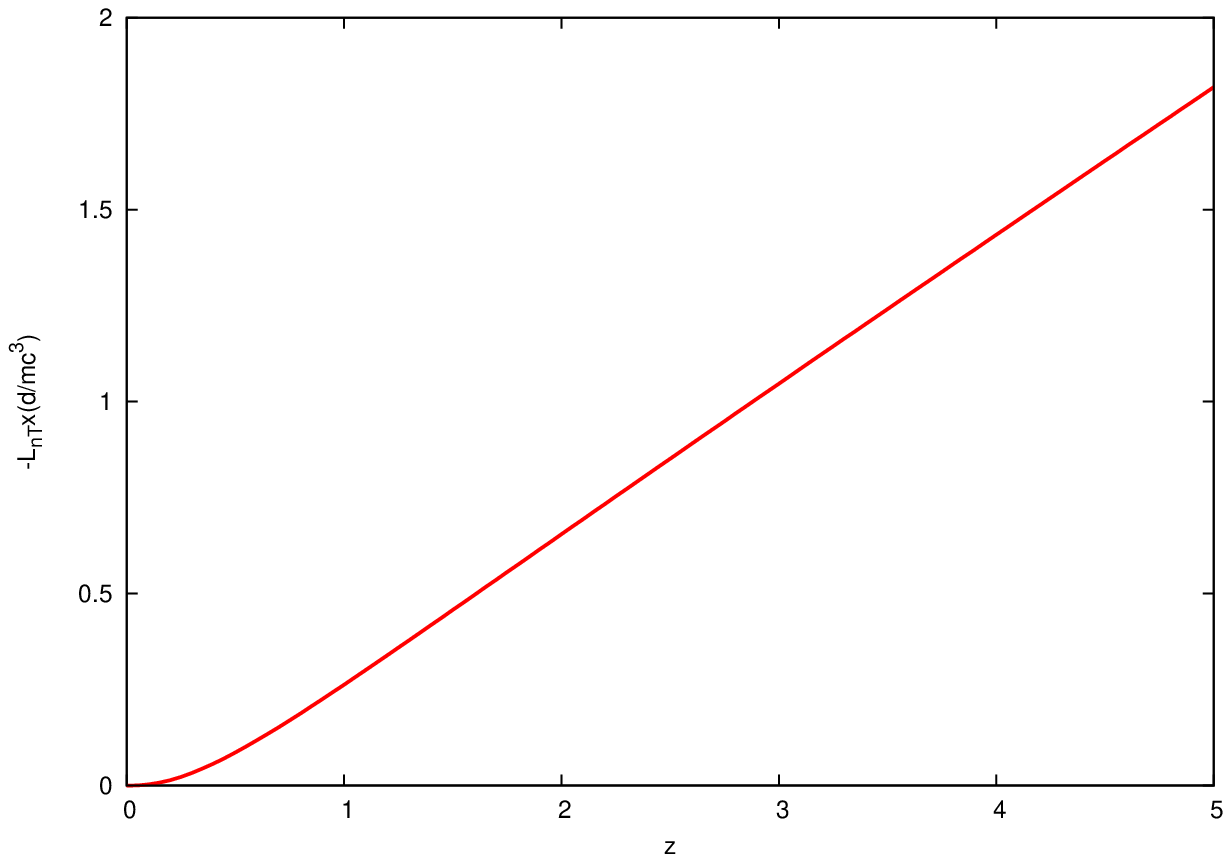}
\caption{\label{fig:2}The dimensionless bulk viscosity $\mu$ and the coefficient
$L_{nT}$ as functions of the relativistic parameter $z.$}
\end{figure}

\begin{figure}
\includegraphics[scale=0.6]{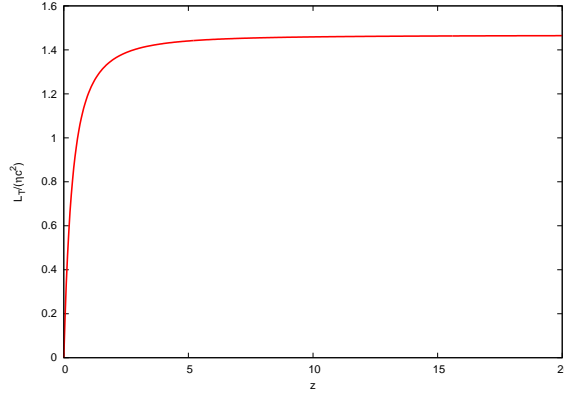}
\caption{\label{fig:3}The ratio $L_T/\left(\eta c^2 \right)$ approaching a constant value
(in dimensionless variables) for large values of $z$.}
\end{figure}

Qualitatively one can clearly observe that the non-relativistic limits
are satisfied with, $\mu/\eta{}_{\left(NR\right)}\rightarrow0$, $\eta/\eta{}_{\left(NR\right)}\rightarrow0$,
$L_{T}/L_{T\left(NR\right)}\rightarrow1$ and $L_{nT}/L_{T\left(NR\right)}\rightarrow0$.
Moreover, since in the limit $z\rightarrow0$, one can approximate
\[
\mathcal{I}_{2}\sim\int_{2/z}^{\infty}e^{-\left(x-\frac{2}{z}\right)}x\left(z^{2}x^{2}-4\right)^{5/2}dx=
240e^{\frac{2}{z}}z\mathcal{K}_{4}\left(\frac{1}{z}\right),
\]
\[
\mathcal{I}_{3}\sim\int_{2/z}^{\infty}e^{-\left(x-\frac{2}{z}\right)}\left(z^{2}x^{2}-4\right)^{5/2}dx=120e^{\frac{2}{z}}z^{2}
\mathcal{K}_{3}\left(\frac{1}{z}\right),
\]
where $\mathcal{K}_{n}$ denotes the $n-th$ order modified Bessel
function of the second kind, one has
\[
L_{T}\sim L_{T\left(\text{NR}\right)}\left(1-\frac{105}{48}z+\frac{4515}{1536}z^{2}....\right),
\]
\[
\eta\sim\eta_{\left(\text{NR}\right)}\left(1-\frac{63}{16}z+\frac{4473}{512}z^{2}....\right).
\]
The vanishing at mild and low temperatures of $\mu$ and $L_{nT}$
for the single component monoatomic gas is a known fact which can
be corroborated for example in Ref. \cite{Ch-E}. In Fig. (\ref{fig:2})
one can verify that these coefficients vanish by themselves thus recovering
the Fourier and Navier-Newton relations for these kind of systems
at the low temperature limits. The rigorous examination of these limits
as well as the analysis of the relation of these coefficients with
those obtained by Miller et. al. \cite{millerUR} using a relaxation
approximation and the method of moments deserves a separate discussion
which will be published elsewhere. However, it is worthwhile to notice
at this stage that the bulk viscosity tends to zero and the ratio
$L_{T}/\eta$ approaches a constant value (in dimensionless variables)
for large values of $z$, in agreement with the general results in
such work (see Fig. (\ref{fig:3})). 

\appendix
\section{Appendix A}\label{Appendix A}
In this appendix, the establishment of the expressions for the integrals
of the form $\mathcal{T}_{n}^{\alpha_{1}...\alpha_{n}}=\int f^{\left(0\right)}A\left(\gamma\right)v^{\alpha_{1}}v^{\alpha_{2}}...v^{\alpha_{n}}dv^{*}$,
for any integrable function $A\left(\gamma\right)$, is shown to some
detail for $n=1-4$. These results are used throughout the main text.
The procedure in all cases consists in expressing the molecular velocity
as $v^{\alpha}=v^{\beta}h_{\beta}^{\alpha}+\gamma u^{\alpha}$ together
with the transformation equation Ref. \cite{Micro}
\begin{equation}
v^{\alpha}=\mathcal{L}^{\alpha\beta}K_{\beta}.\label{A1}
\end{equation}
Here $\mathcal{L}^{\alpha\beta}$ is a Lorentz
boost to a system moving with velocity $\vec{w}$, and thus with the
fluid element, and $K_{\beta}$ is the molecule's four velocity in
such a frame, namely the chaotic velocity. One then has, for $n=1$:
\begin{equation}
\int A\left(\gamma\right)v^{\alpha}dv^{*}=\int A\left(\gamma\right)h_{\beta}^{\alpha}v^{\beta}dv^{*}+2\pi c^{2}u^{\alpha}
\int A\left(\gamma\right)\gamma d\gamma,\label{A2}
\end{equation}
and since
\begin{equation}
\int A\left(\gamma\right)h_{\beta}^{\alpha}v^{\beta}dv^{*}=h_{\beta}^{\alpha}\mathcal{L}_{\mu}^{\beta}\int A\left(\gamma\right)
K^{\mu}dv^{*}=h_{\beta}^{\alpha}\mathcal{L}_{i}^{\beta}\int A\left(\gamma\right)K^{i}dv^{*}=0,\label{A3}
\end{equation}
thus
\begin{equation}
\int A\left(\gamma\right)v^{\alpha}dv^{*}=2\pi c^{2}u^{\alpha}\int A\left(\gamma\right)\gamma d\gamma.\label{A4}
\end{equation}
Similarly, for $n=2$,
\begin{eqnarray}
\int A\left(\gamma\right)v^{\mu}v^{\nu}dv^{*} & =u^{\nu}u^{\mu}\int\gamma^{2}A\left(\gamma\right)dv^{*}+u^{\mu}h_{\delta}^{\nu}\int\gamma A\left(\gamma\right)v^{\delta}dv^{*}\nonumber \\
 & +u^{\nu}h_{\sigma}^{\mu}\int\gamma A\left(\gamma\right)v^{\sigma}dv^{*}+\int A\left(\gamma\right)
 \left(h_{\delta}^{\nu}h_{\sigma}^{\mu}v^{\delta}v^{\sigma}\right)dv^{*},
\end{eqnarray}
where the second and third terms vanish in view of Eq. (\ref{A4}).
For the last term we use Eq. (\ref{A1}), by means of which one obtains
\begin{eqnarray}
\int A\left(\gamma\right)\left(h_{\delta}^{\nu}h_{\sigma}^{\mu}v^{\delta}v^{\sigma}\right)dv^{*} & =h_{\delta}^{\nu}h_{\sigma}^{\mu}\mathcal{L}_{i}^{\sigma}\mathcal{L}^{\delta j}\int A\left(\gamma\right)\left(K^{i}K_{j}\right)dv^{*}\nonumber \\
 & =-h_{\delta}^{\nu}h_{\sigma}^{\mu}\mathcal{L}_{i}^{\sigma}\mathcal{L}^{\delta j}\delta_{j}^{i}\int A\left(\gamma\right)
 \gamma^{2}\frac{k^{2}}{2}dv^{*},
\end{eqnarray}
where, for the second equality, we have used the isotropy in the phase
space integral to write
\begin{equation}
\int A\left(\gamma\right)\gamma^{2}\left(k^{1}k_{1}\right)dk^{*}=\int A\left(\gamma\right)\gamma^{2}\left(k^{2}k_{2}\right)
dk^{*}=-\frac{1}{2}\int A\left(\gamma\right)\gamma^{2}k^{2}dk^{*}.\label{A7}
\end{equation}
Finally, since $\gamma^2k^{2}=-c^{2}\left(1-\gamma^{2}\right)$ 
\begin{equation}
\int A\left(\gamma\right)v^{\mu}v^{\nu}dv^{*}=2\pi c^{4}\left\{ \frac{u^{\mu}u^{\nu}}{c^{2}}\int\gamma^{2}A\left(\gamma\right)d\gamma+\frac{1}{2}h^{\mu\nu}\int\left(1-\gamma^{2}\right)A\left(\gamma\right)d\gamma\right\} .\label{A8}
\end{equation}
The integral corresponding to $n=3$ can be reduced to
\begin{eqnarray}
\int A\left(\gamma\right)v^{\mu}v^{\nu}v^{\sigma}dv^{*} & =&2\pi c^{2}u^{\mu}u^{\nu}u^{\sigma}\int A\left(\gamma\right)\gamma^{3}d\gamma+h_{\alpha}^{\mu}h_{\beta}^{\nu}h_{\delta}^{\sigma}\int A\left(\gamma\right)v^{\alpha}v^{\beta}v^{\delta}dv^{*}\nonumber \\
 & + & u^{\sigma}h_{\alpha}^{\mu}h_{\beta}^{\nu}\int A\left(\gamma\right)\gamma v^{\alpha}v^{\beta}dv^{*}+u^{\nu}h_{\alpha}^{\mu}h_{\delta}^{\sigma}\int A\left(\gamma\right)\gamma v^{\alpha}v^{\delta}dv^{*}\nonumber \\
 & + & u^{\mu}h_{\beta}^{\nu}h_{\delta}^{\sigma}\int A\left(\gamma\right)\gamma v^{\delta}v^{\beta}dv^{*},\label{A9}
\end{eqnarray}
since various terms featuring $v^{\alpha}$ in the integrand vanish
as in the previous case. The second term in Eq. (\ref{A9}) also vanishes since
\begin{equation}
h_{\alpha}^{\mu}h_{\beta}^{\nu}h_{\delta}^{\sigma}\int A\left(\gamma\right)v^{\alpha}v^{\beta}v^{\delta}dv^{*}=
h_{\alpha}^{\mu}h_{\beta}^{\nu}h_{\delta}^{\sigma}\mathcal{L}_{\lambda}^{\alpha}\mathcal{L}_{\epsilon}^{\beta}
\mathcal{L}_{\chi}^{\delta}\int A\left(\gamma\right)\gamma^{3}k^{\lambda}k^{\epsilon}k^{\chi}dv^{*}=0,\label{A10}
\end{equation}
and for the rest of the integrals one uses Eq. (\ref{A8}), leading
to
\begin{eqnarray}
u^{\sigma}h_{\alpha}^{\mu}h_{\beta}^{\nu}\int A\left(\gamma\right)\gamma v^{\alpha}v^{\beta}dv^{*} & = & 2\pi c^{4}u^{\sigma}h_{\alpha}^{\mu}h_{\beta}^{\nu}\int A\left(\gamma\right)\left[\frac{u^{\alpha}u^{\beta}}{c^{2}}\gamma^{3}+\frac{1}{2}h^{\alpha\beta}\gamma\left(1-\gamma^{2}\right)\right]d\gamma\nonumber \\
 & = & 2\pi c^{4}\frac{1}{2}u^{\sigma}h^{\mu\nu}\int A\left(\gamma\right)\gamma\left(1-\gamma^{2}\right)d\gamma,
\end{eqnarray}
and analogous expressions for the remaining two terms. Using such
results one is led to the following expression
\begin{eqnarray}
\int A\left(\gamma\right)v^{\mu}v^{\nu}v^{\sigma}dv^{*} & = & 2\pi c^{5}\left\{ \frac{u^{\mu}u^{\nu}u^{\sigma}}{c^{3}}\int A\left(\gamma\right)\gamma^{3}d\gamma\right.\nonumber \\
 &  & \left.+\frac{1}{2c}\left(h^{\mu\nu}u^{\sigma}+h^{\mu\sigma}u^{\nu}+h^{\nu\sigma}u^{\mu}\right)\int A\left(\gamma\right)\gamma
 \left(1-\gamma^{2}\right)d\gamma\right\}.
\end{eqnarray}
Finally, in order to address the case $n=4$, we use Eq. (\ref{A10})
which reduces the integral to the following three terms
\begin{eqnarray}
\int A\left(\gamma\right)v^{\mu}v^{\nu}v^{\sigma}v^{\lambda}dv^{*} & = & 2\pi c^{2}u^{\mu}u^{\nu}u^{\sigma}u^{\lambda}\int A\left(\gamma\right)\gamma^{4}d\gamma+h_{\alpha}^{\mu}h_{\beta}^{\nu}h_{\delta}^{\sigma}h_{\epsilon}^{\lambda}\int A\left(\gamma\right)v^{\alpha}v^{\beta}v^{\delta}v^{\epsilon}dv^{*}\nonumber \\
 & + &\pi c^{4}\left(u^{\sigma}u^{\lambda}h^{\mu\nu}+u^{\nu}u^{\lambda}h^{\mu\sigma}+u^{\mu}u^{\lambda}h^{\nu\sigma}+u^{\sigma}u^{\nu}h^{\mu\lambda}\right.\nonumber \\
 & + &  \left. u^{\sigma}u^{\mu}h^{\nu\lambda}+u^{\mu}u^{\nu}h^{\lambda\sigma}\right)\int A\left(\gamma\right)\gamma^{2}\left(1-\gamma^{2}\right)d\gamma.
\end{eqnarray}
For the second term, one can write
\begin{equation}
h_{\alpha}^{\mu}h_{\beta}^{\nu}h_{\delta}^{\sigma}h_{\epsilon}^{\lambda}\int A\left(\gamma\right)v^{\alpha}v^{\beta}v^{\delta}v^{\epsilon}dv^{*}=h_{\alpha}^{\mu}h_{\beta}^{\nu}h_{\delta}^{\sigma}h_{\epsilon}^{\lambda}\mathcal{L}_{a}^{\alpha}\mathcal{L}_{b}^{\beta}\mathcal{L}_{c}^{\delta}\mathcal{L}_{d}^{\epsilon}\int\gamma^{7}A\left(\gamma\right)k^{a}k^{b}k^{c}k^{d}d^{2}k
\end{equation}
with $a,\,b,\,c,\,d=1,2$ and $d^{2}k=kdkd\theta$. Integrating over
$\theta$ and using $\mathcal{L}_{i}^{\sigma}\mathcal{L}^{\delta j}\delta_{j}^{i}=\mathcal{L}_{i}^{\sigma}\mathcal{L}^{\delta i}=h^{\sigma\delta}$
yields
\begin{equation}
h_{\alpha}^{\mu}h_{\beta}^{\nu}h_{\delta}^{\sigma}h_{\epsilon}^{\lambda}\int A\left(\gamma\right)v^{\alpha}v^{\beta}v^{\delta}
v^{\epsilon}dv^{*}=\frac{\pi c^{6}}{4}\left(h^{\mu\nu}h^{\sigma\lambda}+h^{\mu\lambda}h^{\nu\sigma}+h^{\mu\sigma}h^{\nu\lambda}
\right)\int A\left(\gamma\right)\left(\gamma^{2}-1\right)^{2}d\gamma,
\end{equation}
and thus
\begin{eqnarray}
\int A\left(\gamma\right)v^{\mu}v^{\nu}v^{\sigma}v^{\lambda}dv^{*} & = &\pi c^{6}\left\{ \frac{1}{4}\left(h^{\mu\nu}h^{\sigma\lambda}
+h^{\mu\lambda}h^{\nu\sigma}+h^{\mu\sigma}h^{\nu\lambda}\right)\int A\left(\gamma\right)\left(\gamma^{2}-1\right)^{2}d\gamma\right.\nonumber \\
& + & \frac{1}{c^{2}} \left(u^{\sigma}u^{\lambda}h^{\mu\nu}+u^{\nu}u^{\lambda}h^{\mu\sigma}+u^{\mu}u^{\lambda}h^{\nu\sigma}+u^{\sigma}u^{\nu}h^{\mu\lambda}+u^{\sigma}u^{\mu}h^{\nu\lambda}+u^{\mu}u^{\nu}h^{\lambda\sigma}\right)\nonumber \\
& \times  & \left.\int\gamma^{2}\left(1-\gamma^{2}\right)A\left(\gamma\right)d\gamma+\frac{2}{c^{4}}u^{\mu}u^{\nu}u^{\sigma}u^{\lambda}\int\gamma^{4}A\left(\gamma\right)d\gamma\right\} .
\end{eqnarray}
\appendix
\section{Appendix B} \label{Appendix B}
In order to reduce the proposed general solution given in Eq. (\ref{30})
to the one stated in Eq. (\ref{33}), one starts by writing
$\mathcal{A}_{\alpha\beta}$ in a general way, in terms of all the symmetric second rank
tensors
\begin{equation}
\mathcal{A}_{\alpha\beta}=a\eta_{\alpha\beta}+bu_{\alpha}u_{\beta}+d\left(v_{\alpha}u_{\beta}+v_{\beta}u_{\alpha}\right)
+ev_{\alpha}v_{\beta},\label{B1}
\end{equation}
since any antisymmetrical component would vanish upon contraction with
$\mathring{\sigma}^{\alpha\beta}$. Also, since $\mathring{\sigma}^{\alpha\beta}$
is orthogonal to both $u^{\alpha}$ and $u^{\beta}$, $b$ and $d$
can be chosen to be zero. Similarly one can consider $a=0$ since
$\mathring{\sigma}^{\alpha\beta}\eta_{\alpha\beta}=0$. Thus, by
renaming $e=A_{4}\left(\gamma_{k}\right)$, the term in Eq. (\ref{30}) corresponding
to the traceless symmetric part of the velocity gradient is given
solely by $A_{4}\left(\gamma_{k}\right)v_{\alpha}v_{\beta}\mathring{\sigma}^{\alpha\beta}$.

In order to impose the subsidiary conditions, we start by rewriting
Eq. (\ref{31}) as
{\small 
\begin{equation}
\int\left[\left(\mathcal{A}_{1}\left(\gamma_{k}\right)\frac{T_{,\alpha}}{T}+\mathcal{A}_{2}\left(\gamma_{k}\right)\frac{n_{,\alpha}}{n}\right)v^{\beta}h_{\beta}^{\alpha}
+\mathcal{A}_{3}\left(\gamma_{k}\right)u_{,\alpha}^{\alpha}\right.
\left.+A_{4}\left(\gamma_{k}\right)v_{\alpha}v_{\beta}\mathring{\sigma}^{\alpha\beta}+\alpha+\check{\alpha}_{\alpha}v^{\alpha}\right]\gamma_{k}^{2}f^{(0)}d\gamma_{k}=0.\label{B2}
\end{equation}}
Using the integrals (\ref{A3}) and (\ref{A8}) given
in Appendix A and since $\mathring{\sigma}^{\alpha\beta}h_{\alpha\beta}=0$,
the previous equation can be reduced to

\begin{equation}
\int\left[\left(\mathcal{A}_{3}\left(\gamma_k\right)u_{,\alpha}^{\alpha}+\alpha\right)\gamma_{k}^{2}+\check{\alpha}_{\alpha}u^{\alpha}\gamma_{k}^{3}\right]f^{\left(0\right)}d\gamma_{k}=0.\label{B3}
\end{equation}
Similarly, using again Appendix A, one can show that Eq. (\ref{32})
can be written as
\begin{eqnarray}
\int\left\{ c^{2}h^{\mu\alpha}\left(1-\gamma_{k}^{2}\right)\left(\mathcal{A}_{1}
\left(\gamma_{k}\right)\frac{T_{,\alpha}}{T}+\mathcal{A}_{2}\left(\gamma_{k}\right)
\frac{n_{,\alpha}}{n}+\check{\alpha}_{\alpha}\right) \right. \\\nonumber
 \left. +2u^{\mu} \left[\left(\mathcal{A}_{3}\left(\gamma_{k}
\right)u_{,\alpha}^{\alpha}+\alpha\right)\gamma_{k}+\check{\alpha}_{\alpha}u^{\alpha}\gamma_{k}^{2}\right]
 \right\} &  & f^{\left(0\right)}d\gamma_{k}=0.\label{B5}
\end{eqnarray}
Separating the parallel and orthogonal components with respect to $u^{\mu}$ one obtains
two independent conditions namely,
\begin{equation}
h^{\mu\alpha}\int f^{\left(0\right)}\left(1-\gamma_{k}^{2}\right)\left(\mathcal{A}_{1}\left(\gamma_{k}\right)\frac{T_{,\alpha}}{T}
+\mathcal{A}_{2}\left(\gamma_{k}\right)\frac{n_{,\alpha}}{n}+\check{\alpha}_{\alpha}\right)d\gamma_{k}=0,\label{B6}
\end{equation}
and
\begin{equation}
\int f^{\left(0\right)}\left[\left(\mathcal{A}_{3}\left(\gamma_{k}\right)u_{,\alpha}^{\alpha}+\alpha\right)\gamma_{k}+
\check{\alpha}_{\alpha}u^{\alpha}\gamma_{k}^{2}\right]d\gamma_{k}=0.\label{B7}
\end{equation}
From Eqs. (\ref{B3}) and (\ref{B7}) one can readily deduce that
both $\alpha$ and$\check{\alpha}_{\alpha}u^{\alpha}$ are proportional
to $u_{,\alpha}^{\alpha}$ and thus can be omitted. On the other
hand, $h^{\mu\alpha}\check{\alpha}_{\alpha}$ can also be omitted,
by means of Eq. (\ref{B6}), and thus the solution to be proposed
can be simplified in order to read
\begin{equation}
\phi\left(v^{\nu}\right)=\left(A_{1}\left(\gamma_{k}\right)\frac{T_{,\alpha}}{T}+A_{2}\left(\gamma_{k}\right)\frac{n_{,\alpha}}{n}
\right)v^{\beta}h_{\beta}^{\alpha}+A_{3}\left(\gamma_{k}\right)u_{,\alpha}^{\alpha}+A_{4}\left(\gamma_{k}\right)v_{\alpha}v_{\beta}
\mathring{\sigma}^{\alpha\beta},\label{B8}
\end{equation}
and the subsidiary conditions are reduced to
\begin{equation}
h^{\mu\alpha}\int\left(1-\gamma_{k}^{2}\right)\left(A_{1}\left(\gamma_{k}\right)\frac{T_{,\alpha}}{T}+A_{2}\left(\gamma_{k}\right)
\frac{n_{,\alpha}}{n}\right)f^{\left(0\right)}d\gamma_{k}=0,\label{B9}
\end{equation}
and
\begin{equation}
\int A_{3}\left(\gamma\right)\gamma_{k}f^{\left(0\right)}d\gamma_{k}=\int A_{3}\left(\gamma_{k}\right)\gamma_{k}^{2}f^{\left(0\right)}d\gamma_{k}=0.\label{B10}
\end{equation}
\appendix
\section{Appendix C}\label{Appendix C}
As mentioned above, the values for the coefficients $\alpha_{nj}^{(i)}$ required
in this work are obtained by a standard Gram-Schmidt procedure with
the weight functions $p^{\left(1\right)}\left(z\right)=p^{\left(2\right)}\left(z\right)=e^{-\frac{\gamma}{z}}\left(1-\gamma^{2}\right)$,
$p^{\left(3\right)}\left(z\right)=e^{-\frac{\gamma}{z}}$ and $p^{\left(4\right)}\left(z\right)=e^{-\frac{\gamma}{z}}\left(1-\gamma^{2}\right)^{2}$.
By performing the corresponding calculations one can directly obtain
the following expressions
\begin{equation}
\alpha_{11}^{\left(1\right)}=\alpha_{11}^{\left(2\right)}=\sqrt{\frac{\left(z+1\right)e^{1/z}}{2z^{4}\left(3z^{2}+6z+2\right)}},\label{C1}
\end{equation}
\begin{equation}
\alpha_{10}^{(1)}=\alpha_{10}^{(2)}=-\sqrt{\frac{\left(z+1\right)e^{1/z}}{2z^{4}\left(3z^{2}+6z+2\right)}}\frac{1+3z+2z^{2}}{1+z},\label{eq:C1bis}
\end{equation}
\begin{equation}
\alpha_{00}^{\left(3\right)}=\sqrt{\frac{e^{1/z}}{z}},\label{C2}
\end{equation}
\begin{equation}
\alpha_{10}^{\left(3\right)}=-\sqrt{\frac{e^{1/z}}{z}}\frac{\left(1+z\right)}{z},\label{C3}
\end{equation}
\begin{equation}
\alpha_{11}^{\left(3\right)}=\sqrt{\frac{e^{1/z}}{z}}\frac{1}{z},\label{C4}
\end{equation}
\begin{equation}
\alpha_{20}^{\left(3\right)}=\sqrt{\frac{e^{1/z}}{z}}\frac{1}{2z^{2}}\left(2z^{2}+4z+1\right),\label{C5}
\end{equation}
\begin{equation}
\alpha_{21}^{\left(3\right)}=-\sqrt{\frac{e^{1/z}}{z}}\frac{\left(1+2z\right)}{z^{2}},\label{C6}
\end{equation}
\begin{equation}
\alpha_{22}^{\left(3\right)}=\sqrt{\frac{e^{1/z}}{z}}\frac{1}{2z^{2}},\label{C7}
\end{equation}
\begin{equation}
\alpha_{00}^{\left(4\right)}=\sqrt{\frac{e^{1/z}}{8z^{2}\left(3z^{2}+3z+1\right)}}.\label{C8}
\end{equation}
\appendix
\section{Appendix D}\label{Appendix D}
In this appendix, the collision brackets $\left[\gamma^{2},\gamma^{2}\right]$,
$\left[\gamma v^{\alpha},\gamma v_{\alpha}\right]$ and $\left[v^{\alpha}v^{\beta},v_{\alpha}v_{\beta}\right]$
are calculated to some detail. Starting from the general definition
\begin{equation}
\left[G,H\right]=-\frac{1}{n^{2}}\int G\left[H'_{1}+H'-H_{1}-H\right]f^{\left(0\right)}f_{1}^{\left(0\right)}F\sigma\left(Q,\chi\right)d\chi dv_{1}^{*}dv^{*},\label{eq:general}
\end{equation}
the brackets can be written as

\begin{equation}
\left[\gamma^{2},\gamma^{2}\right]=-\frac{1}{n^{2}}\int f^{\left(0\right)}f_{1}^{\left(0\right)}\gamma^{2}\left(\gamma{}_{1}^{'2}+\gamma{}^{'2}-\gamma_{1}^{2}-\gamma{}^{2}\right)F\sigma\left(Q,\chi\right)d\chi dv^{*}dv_{1}^{*},\label{eq:corchete1}
\end{equation}
\begin{equation}
\left[\gamma v_{\beta},\gamma v^{\beta}\right]=-\frac{1}{n^{2}}\int f^{\left(0\right)}f_{1}^{\left(0\right)}\gamma v_{\beta}
\left(\gamma{}_{1}^{'}v_{1}^{'\beta}+\gamma^{'}v^{'\beta}-\gamma{}_{1}v_{1}^{\beta}-\gamma v^{\beta}\right)F\sigma
\left(Q,\chi\right)d\chi dv^{*}dv_{1}^{*},\label{eq:corchete2}
\end{equation}
\begin{equation}
\left[v_{\alpha}v_{\beta},v^{\alpha}v^{\beta}\right]=-\frac{1}{n^{2}}\int f^{\left(0\right)}f_{1}^{\left(0\right)}v_{\alpha}v_{\beta}\left[v_{1}^{'\alpha}v{}_{1}^{'\beta}+v{}^{'\alpha}v{}^{'\beta}-v{}_{1}^{\alpha}v{}_{1}^{\beta}-v{}^{\alpha}v{}^{\beta}\right]F\sigma\left(Q,\chi\right)d\chi dv^{*}dv_{1}^{*}.\label{eq:corchete3}
\end{equation}
Introducing the change of variables
\[
P^{\alpha}=v^{\alpha}+v_{1}^{\alpha}\qquad Q^{\alpha}=v^{\alpha}-v_{1}^{\alpha},
\]
momentum conservation implies
\[
P^{\alpha'}=P^{\alpha},
\]
and choosing the center of mass frame, where $\left(P^{\alpha}\right)=\left(P^{0},0,0\right)$,
one can rewrite Eqs. (\ref{eq:corchete1}-\ref{eq:corchete3}) as
follows
\begin{equation}
\left[\gamma^{2},\gamma^{2}\right]=-\frac{u_{\mu}u_{\beta}u_{\alpha}u_{\nu}}{16c^{7}n^{2}}\int f^{\left(0\right)}f_{1}^{\left(0\right)}
\left(P^{\mu}P^{\beta}+P^{\mu}Q^{\beta}+Q^{\mu}P^{\beta}+Q^{\mu}Q^{\beta}\right)
\left(Q'^{\nu}Q'^{\alpha}-Q^{\alpha}Q^{\nu}\right)d \vartheta,\label{eq:corchete1-2}
\end{equation}
\begin{equation}
\left[\gamma v_{\beta},\gamma v^{\beta}\right]=-\frac{u_{\alpha}u_{\mu}}{16c^{3}n^{2}}\int f^{\left(0\right)}f_{1}^{\left(0\right)}
\left(P^{\mu}P_{\beta}+P^{\mu}Q_{\beta}+Q^{\mu}P_{\beta}+Q^{\mu}Q_{\beta}\right)
\left(Q'^{\beta}Q'^{\alpha}-Q^{\beta}Q^{\alpha}\right)d \vartheta,\label{eq:corchete2-2}
\end{equation}
\begin{equation}
\left[v_{\alpha}v_{\beta},v^{\alpha}v^{\beta}\right]=-\frac{c}{16n^{2}}\int f^{\left(0\right)}f_{1}^{\left(0\right)}\left(P_{\alpha}P_{\beta}
+P_{\alpha}Q_{\beta}+Q_{\alpha}P_{\beta}+Q_{\alpha}Q_{\beta}\right)\left(Q'^{\beta}Q'^{\alpha}-Q^{\beta}Q^{\alpha}\right)d \vartheta,\label{eq:corchete3-2}
\end{equation}
where $d \vartheta=\sigma\left(\chi\right)d \chi \left(Q/P^0\right)d^{2}Pd^{2}Q$ and $Q^{\alpha}Q_{\alpha}=-Q^{2}$. The scattering cross section for hard disks is given by
\[
\sigma\left(\chi\right)=\frac{d}{2}\left|\sin\left(\frac{\chi}{2}\right)\right|,
\]
where $\chi$ represents the scattering angle. If the
$x_{2}$ axis is aligned in the $Q^{\alpha}$ direction one has
\[
Q^{\alpha}=\left(\begin{array}{c}
0\\
0\\
1
\end{array}\right)\qquad Q^{'\alpha}=\left(\begin{array}{c}
0\\
\sin\chi\\
\cos\chi
\end{array}\right),
\]
which leads to
\[
\int_{0}^{2\pi}\left(Q^{'\mu}Q^{'\beta}-Q^{\mu}Q^{\beta}\right)\sigma\left(\chi\right)d\chi=\frac{16d}{15}Q^{2}
\left[\frac{P^{\mu}P^{\beta}}{P^{2}}-\eta^{\mu\beta}-2\frac{Q^{\mu}Q^{\beta}}{Q^{2}}\right],
\]
where $d$ is the diameter of the disks. Introducing this result in
Eqs. (\ref{eq:corchete1-2}-\ref{eq:corchete3-2}) yields
\begin{equation}
\left[\gamma^{2},\gamma^{2}\right]=-\frac{du_{\mu}u_{\beta}u_{\alpha}u_{\nu}}{15c^{7}n^{2}}\int f^{\left(0\right)}f_{1}^{\left(0\right)}\left(P^{\mu}P^{\beta}
+P^{\mu}Q^{\beta}+Q^{\mu}P^{\beta}+Q^{\mu}Q^{\beta}\right)\chi^{\nu\alpha}
\frac{Q^{3}}{P^{0}}d^{2}Pd^{2}Q,\label{eq:corchete1-3}
\end{equation}
\begin{equation}
\left[\gamma v_{\beta},\gamma v^{\beta}\right]=-\frac{du_{\alpha}u_{\mu}}{15c^{3}n^{2}}\int f^{\left(0\right)}f_{1}^{\left(0\right)}\left(P^{\mu}P_{\beta}
+P^{\mu}Q_{\beta}+Q^{\mu}P_{\beta}+Q^{\mu}Q_{\beta}\right)\chi^{\beta\alpha}
\frac{Q^{3}}{P^{0}}d^{2}Pd^{2}Q,\label{eq:corchete2-3}
\end{equation}
\begin{equation}
\left[v_{\alpha}v_{\beta},v^{\alpha}v^{\beta}\right]=-\frac{dc}{15n^{2}}\int f^{\left(0\right)}f_{1}^{\left(0\right)}\left(P_{\alpha}P_{\beta}+P_{\alpha}Q_{\beta}
+Q_{\alpha}P_{\beta}+Q_{\alpha}Q_{\beta}\right)\chi^{\beta\alpha}
\frac{Q^{3}}{P^{0}}d^{2}Pd^{2}Q,\label{eq:corchete3-3}
\end{equation}
where we have defined
\begin{equation}
 \chi^{\nu\alpha}=\left[\frac{P^{\nu}P^{\alpha}}{P^{2}}-\eta^{\nu\alpha}-2\frac{Q^{\nu}Q^{\alpha}}{Q^{2}}\right].
\end{equation}
In order to calculate the integrals in the polar angle we use
\[
\left(Q^{\alpha}\right)=Q\left(\begin{array}{c}
0\\
\cos\phi\\
\sin\phi
\end{array}\right),\qquad d^{2}Q=Qd\phi dQ
\]
and after a somewhat lengthy procedure, one can obtain
\begin{equation}
\left[\gamma^{2},\gamma^{2}\right]=\frac{\pi d \zeta\left(z\right)^{2}}{30c^{7}}\int e^{-\frac{u_{\alpha}P^{\alpha}}{zc^{2}}}u_{\alpha}u_{\beta}
\left[u_{\mu}u_{\nu}\frac{P^{\alpha}P^{\beta}P^{\mu}P^{\nu}}{P^{4}}-2c^{2}\frac{P^{\alpha}P^{\beta}}{P^{2}}+c^{2}\eta^{\alpha \beta}
\right]\frac{Q^{6}}{P^{0}}d^{2}PdQ,\label{eq:corchete1-4}
\end{equation}

\begin{equation}
\left[\gamma v_{\beta},\gamma v^{\beta}\right]=-\frac{\pi d \zeta\left(z\right)^{2}}{15c^{3}}\int e^{-\frac{u_{\alpha}P^{\alpha}}{zc^{2}}}u_{\mu}u_{\alpha}
\left[\frac{P^{\alpha}P^{\mu}}{P^{2}}-\eta^{\alpha\mu}\right]\frac{Q^{6}}{P^{0}}d^{2}PdQ,\label{eq:corchete2-4}
\end{equation}

\begin{equation}
\left[v_{\alpha}v_{\beta},v^{\alpha}v^{\beta}\right]=\frac{2\pi d\zeta\left(z\right)^{2}}{15c^{-1}}\int e^{-\frac{u_{\alpha}
P^{\alpha}}{zc^{2}}}\frac{Q^{6}}{P^{0}}d^{2}PdQ.\label{eq:corchete3-4}
\end{equation}
with $\zeta\left(z\right)=\left(1/2\pi c^2\right)\left(e^{1/z}/\left(z+z^2\right)\right)$.
Now we can identify the following functions in the above equations,
\begin{equation}
Z^{*}=\int e^{-\frac{U_{\alpha}P^{\alpha}}{zc^{2}}}\frac{d^{2}P}{P^{0}},\label{eq:z1}
\end{equation}
\begin{equation}
Z_{\mu\alpha}^{*}=\int e^{-\frac{U_{\alpha}P^{\alpha}}{zc^{2}}}P_{\alpha}P_{\mu}\frac{d^{2}P}{P^{0}},\label{eq:z2}
\end{equation}
\begin{equation}
Z_{\alpha\beta\mu\nu}^{*}=\int e^{-\frac{U_{\alpha}P^{\alpha}}{zc^{2}}}P_{\alpha}P_{\beta}P_{\mu}P_{\nu}\frac{d^{2}P}{P^{0}},\label{eq:z3}
\end{equation}
and rewrite Eqs. (\ref{eq:corchete1-4}-\ref{eq:corchete3-4}) as
\begin{equation}
\left[\gamma^{2},\gamma^{2}\right]=\frac{\pi d}{30c^{7}}\zeta\left(z\right)^{2}\int\left(\frac{u_{\alpha}u_{\beta}u_{\mu}
u_{\nu}}{P^{4}}Z_{\alpha\beta\mu\nu}^{*}-2c^{2}\frac{u_{\alpha}u_{\beta}}{P^{2}}Z_{\alpha\beta}^{*}+c^{4}Z^{*}\right)Q^{6}dQ,\label{eq:corchete1-5}
\end{equation}
\begin{equation}
\left[\gamma v_{\beta},\gamma v^{\beta}\right]=-\frac{d\pi}{15c^{3}}\zeta\left(z\right)^{2}\int\left(\frac{u_{\mu}
u_{\alpha}}{P^{2}}Z_{\mu\alpha}^{*}-c^{2}Z^{*}\right)Q^{6}dQ,\label{eq:corchete2-5}
\end{equation}
\begin{equation}
\left[v_{\alpha}v_{\beta},v^{\alpha}v^{\beta}\right]=\frac{2\pi d}{15c^{-1}}\zeta\left(z\right)^{2}\int Z^{*}Q^{6}dQ.\label{eq:corchete3-5}
\end{equation}

The integrals in Eqs. (\ref{eq:z1}-\ref{eq:z3}) are evaluated in some
detail in Ref. \cite{HFBidi} following the procedure shown in Ref. \cite{CercignaniKremer}, which yield 
\[
Z^{*}=2\pi cz\exp\left(-\frac{\mathcal{Q}}{z}\right),
\]
\[
u^{\mu}u^{\alpha}Z_{\mu\alpha}^{*}=2\pi c^{5}z\exp\left(-\frac{\mathcal{Q}}{z}\right)\left[\mathcal{Q}^{2}+2\mathcal{Q}z+2z^{2}
\right],
\]
\[
u^{\alpha}u^{\beta}u^{\mu}u^{\nu}Z_{\alpha\beta\mu\nu}^{*}=2\pi c^{9}z\exp\left(-\frac{\mathcal{Q}}{z}\right)\left[12\mathcal{Q}^{2}z^{2}+24\mathcal{Q}z^{3}+24z^{4}+4\mathcal{Q}^{3}z+\mathcal{Q}^{4}\right],
\]
where we have also used energy-momentum conservation $P^{2}=4c^{2}+Q^{2}$
and the change of variable $Q^{2}+4c^{2}=c^{2}\mathcal{Q}^{2}$. Thus, one can write Eqs. (\ref{eq:corchete1-5}-\ref{eq:corchete3-5}) 
as follows
\begin{equation}
\left[\gamma^{2},\gamma^{2}\right]=\frac{2czd}{15}\frac{e^{2/z}}{\left(z+z^{2}\right)^{2}}\int_{2}^{\infty}\exp
\left(-\frac{\mathcal{Q}}{z}\right)\left(\frac{z^{2}}{\mathcal{Q}^{2}}+3\frac{z^{3}}{\mathcal{Q}^{3}}
+3\frac{z^{4}}{\mathcal{Q}^{4}}\right)\left(\mathcal{Q}^{2}-4\right)^{5/2}\mathcal{Q}d\mathcal{Q},\label{eq:corchete1-6}
\end{equation}
\begin{equation}
\left[\gamma v_{\beta},\gamma v^{\beta}\right]=-\frac{c^{3}zd}{15}\frac{e^{2/z}}{\left(z+z^{2}\right)^{2}}\int_{2}^{\infty}\exp
\left(-\frac{\mathcal{Q}}{z}\right)\left(\frac{z}{\mathcal{Q}}+\frac{z^{2}}{\mathcal{Q}^{2}}\right)
\left(\mathcal{Q}^{2}-4\right)^{5/2}\mathcal{Q}d\mathcal{Q},\label{eq:corchete2-6}
\end{equation}
\begin{equation}
\left[v_{\alpha}v_{\beta},v^{\alpha}v^{\beta}\right]=\frac{c^{5}zd}{15}\frac{e^{2/z}}{\left(z+z^{2}\right)^{2}}\int_{2}^{\infty}
\exp\left(-\frac{\mathcal{Q}}{z}\right)\left(\mathcal{Q}^{2}-4\right)^{5/2}\mathcal{Q}d\mathcal{Q}.\label{eq:corchete3-6}
\end{equation}
and introducing the change of variable $x=\mathcal{Q}/z$ one obtains
\begin{equation}
\left[\gamma^{2},\gamma^{2}\right]=\frac{2cz^{3}d}{15}\frac{e^{2/z}}{\left(z+z^{2}\right)^{2}}\int_{2/z}^{\infty}e^{-x}
\left(\frac{1}{x}+\frac{3}{x^{2}}+\frac{3}{x^{3}}\right)\left(z^{2}x^{2}-4\right)^{5/2}dx,\label{eq:c1final}
\end{equation}
\begin{equation}
\left[\gamma v_{\beta},\gamma v^{\beta}\right]=-\frac{c^{3}z^{3}d}{15}\frac{e^{2/z}}{\left(z+z^{2}\right)^{2}}
\int_{2/z}^{\infty}e^{-x}\left(1+\frac{1}{x}\right)\left(z^{2}x^{2}-4\right)^{5/2}dx,\label{eq:c2final}
\end{equation}
\begin{equation}
\left[v_{\alpha}v_{\beta},v^{\alpha}v^{\beta}\right]=\frac{c^{5}z^{3}d}{15}\frac{e^{2/z}}{\left(z+z^{2}\right)^{2}}
\int_{2/z}^{\infty}e^{-x}x\left(z^{2}x^{2}-4\right)^{5/2}dx,\label{eq:c3final}
\end{equation}
these integrals can be obtained numerically for a given value of $z$.
\bibliography{Rel}
\bibliographystyle{unsrt}
\end{document}